\documentclass{vldb}
\usepackage{graphicx}
\usepackage{amssymb}
\usepackage{multirow}
\usepackage{subfigure}
\usepackage{amsmath, thm-restate}
\usepackage{algorithmic}
\usepackage[section]{algorithm}
\usepackage{subfigure}
\usepackage{url}

\begin{document}
\toappear{}
\pagenumbering{arabic}

\def \bb{\:\:\blacktriangleright\!\!}

\newcommand{\mh}[1]{[[\emph{MH: #1}]]}
\newcommand{\vr}[1]{[[\emph{VR: #1}]]}
\newcommand{\gm}[1]{[[\emph{GM: #1}]]}
\newcommand{\ds}[1]{[[\emph{ds: #1}]]}
\newcommand{\eat}[1]{}
\newcommand{\cut}[1]{}

\newcommand{\set}[1]{\{#1\}}   

\newtheorem{definition}{Definition}[section]
\newtheorem{proposition}{Proposition}
\newtheorem{conjecture}{Conjecture}
\newtheorem{theorem}{Theorem}
\newtheorem{problem}{Problem}
\newtheorem{example}{Example}

\def\error{error}
\def\nbrs{nbrs}
\def\<{\langle}
\def\>{\rangle}

\newcommand{\sens}[1]{\Delta_{#1}}

\def\Q{\mathbf{Q}}
\def\QQ{\mathbf{\tilde{Q}}}
\def\QC{\mathbf{\overline{Q}}}
\def\q{x}
\def\qq{\tilde{q}}
\def\qbar{\overline{q}}

\def\H{\mathbf{H}}
\def\HH{\mathbf{\tilde{H}}}
\def\HC{\mathbf{\overline{H}}}
\def\hh{\tilde{h}}
\def\hbar{\overline{h}}

\def\L{\mathbf{L}}
\def\LL{\mathbf{\tilde{L}}}
\def\lsmall{\tilde{l}}
\def\LC{\mathbf{\overline{L}}}

\def\S{\mathbf{S}}
\def\SS{\mathbf{\tilde{S}}}
\def\SC{\mathbf{\overline{S}}}
\def\ss{\tilde{s}}
\def\sbar{\overline{s}}

\def\SR{\mathbf{\tilde{S}_r}}
\def\sr{\tilde{s}_r}

\def\Lap{\mbox{Lap}}
\def\cnt{c}
\def\cons{\gamma}
\def\db{I}

\def\hght{{\ell}}  
\def\hv{\hght(v)}
\def\wt{\alpha}
\def\root{r}

\newcommand{\E}{\mathbb{E}}
\newcommand{\Ldist}[3]{||#1 -#2||_{#3}}
\newcommand{\Ltwo}[2]{||#1 -#2||_2}
\newcommand{\Lone}[2]{\left\Vert #1 -#2 \right\Vert_1}

\newtheorem{lemma}{Lemma}

\newcommand{\one}[1]{\mathbb{I}_{#1}}
\def\U{\mathcal U}
\def\Z{succZ} 
\def\s{s}
\def\m{M} 
\def\mm{\tilde{M}}

\title{Boosting the Accuracy of Differentially Private Histograms Through Consistency}

\numberofauthors{2}
\author{
Michael Hay$^\dagger$, Vibhor Rastogi$^\ddagger$, Gerome Miklau$^\dagger$, Dan Suciu$^\ddagger$
\and  
\alignauthor 
	\affaddr{$\dagger$ University of Massachusetts Amherst}\\
	\affaddr{\{mhay,miklau\}@cs.umass.edu}
\alignauthor 
	\affaddr{$\ddagger$ University of Washington}\\
	\affaddr{\{vibhor,suciu\}@cs.washington.edu}
}

\maketitle

\begin{abstract}
We show that it is possible to significantly improve the accuracy of a general class of histogram queries while satisfying differential privacy. Our approach carefully chooses a set of queries to evaluate, and then exploits consistency constraints that should hold over the noisy output. In a post-processing phase, we compute the consistent input most likely to have produced the noisy output.  The final output is differentially-private and consistent, but in addition, it is often much more accurate.  We show, both theoretically and experimentally, that these techniques can be used for estimating the degree sequence of a graph very precisely, and for computing a histogram that can support arbitrary range queries accurately.
\end{abstract}


\section{Introduction}

Recent work in differential privacy \cite{dwork2006calibrating} has shown that it is possible to analyze sensitive data while ensuring strong privacy guarantees.  Differential privacy is typically achieved through random perturbation: the analyst issues a query and receives a noisy answer.  To ensure privacy, the noise is carefully calibrated to the {\em sensitivity} of the query.  Informally, query sensitivity measures how much a small change to the database---such as adding or removing a person's private record---can affect the query answer.  Such query mechanisms are simple, efficient, and often quite accurate.
In fact, one mechanism has recently been shown to be optimal for a single counting query \cite{ghosh2009universally}---i.e., there is no better noisy answer to return under the desired privacy objective.


However, analysts typically need to compute multiple sta\-tistics on a database. 
Differentially private algorithms extend nicely to a set of queries, but there can be difficult trade-offs among alternative strategies for answering a workload of queries.  Consider the analyst of a private student database who requires answers to the following queries: the total number of students, $\q_t$, the number of students $\q_A$, $\q_B$, $\q_C$, $\q_D$, $\q_F$ receiving grades A, B, C, D, and F respectively, and the number of passing students, $\q_p$ (grade D or higher).  

Using a differentially private interface, a first alternative is to request noisy answers for just $(\q_A, \q_B, \q_C, \q_D, \q_F)$ and use those answers to compute answers for $\q_t$ and $\q_p$ by summation.  The sensitivity of this set of queries is 1 because adding or removing one tuple changes exactly one of the five outputs by a value of one. Therefore, the noise added to individual answers is low and the noisy answers are accurate estimates of the truth.  Unfortunately, the noise accumulates under summation, so the estimates for $\q_t$ and $\q_p$ are worse. 
%

A second alternative is to request noisy answers for all queries~$(\q_t, \q_p,$ $\q_A,$ $\q_B, \q_C, \q_D, \q_F)$.  This query set has sensitivity 3 (one change could affect three return values, each by a value of one), and the privacy mechanism must add more noise to each component.  This means the estimates for $\q_A, \q_B,$ $\q_C, \q_D, \q_F$ are worse than above, but the estimates for $\q_t$ and $\q_p$ may be more accurate.  There is another concern, however: inconsistency.  
The noisy answers are likely to violate the following constraints, which one would naturally expect to hold: $\q_t = \q_p + \q_F$ and $\q_p = \q_A + \q_B + \q_C + \q_D$.  This means the analyst must find a way to reconcile the fact that there are two different estimates for the total number of students and two different estimates for the number of passing students.  We propose a technique for resolving inconsistency in a set of noisy answers, and show that doing so can actually increase accuracy.  As a result, we show that strategies inspired by the second alternative can be superior in many cases.

\begin{figure}[t]
\begin{center}
\includegraphics[width=3.25in]{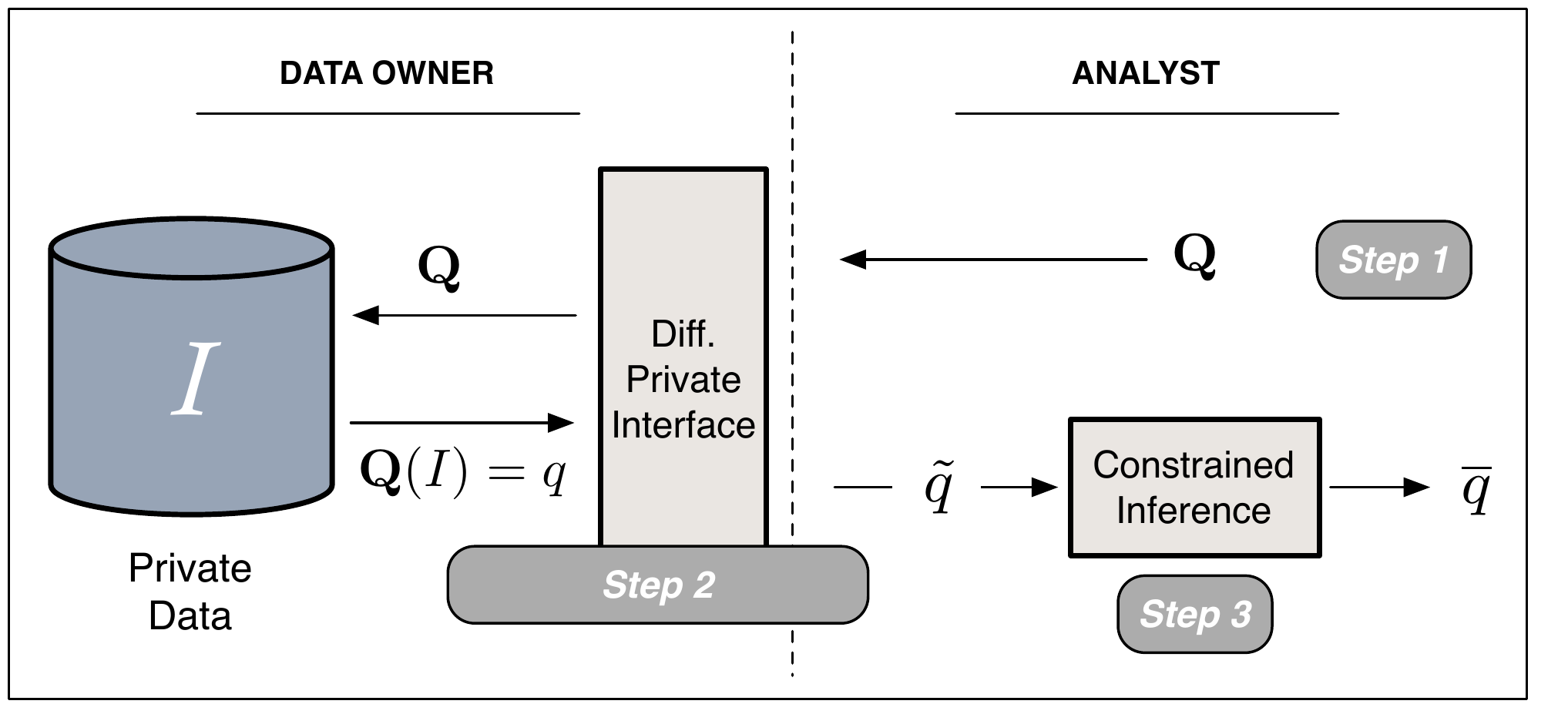}
\caption{
\label{fig:approach} Our approach to querying private data.
}
\end{center}
\vspace{-0.5cm}
\end{figure}

\paragraph*{\bf Overview of Approach} Our approach, shown pictorially in Figure~\ref{fig:approach}, involves three steps.  

\begin{figure*}[!t]
\begin{center}

\subfigure[Trace data]{
\begin{tabular}[b]{c}
	\includegraphics[scale=.5]{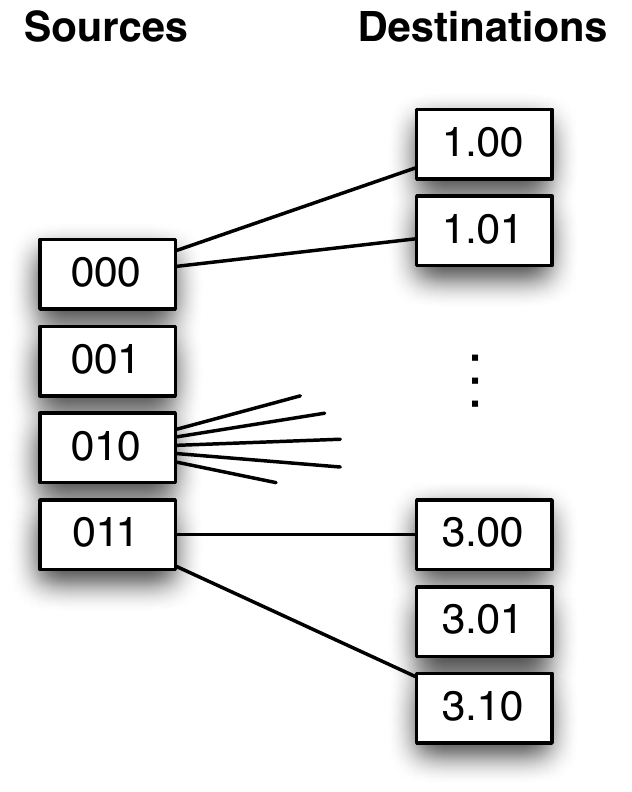}
	\end{tabular}
}
\subfigure[Query variations]{
\begin{tabular}[b]{lll}
Query Definitions: & \multicolumn{2}{l}{$\L:\:\<C_{000},C_{001},C_{010},C_{011}\>$}\\
& \multicolumn{2}{l}{$\H: \<C_{0**}, C_{00*}, C_{01*},C_{000},C_{001},C_{010},C_{011}\>$} \\
& \multicolumn{2}{l}{$\S:~sort(\L)$}\\
\\
{\em True answer}  &  {\em Private output} &  {\em Inferred answer} \\  \hline
$\L(I)=\<2,0,10,2\>$  & $\LL(I) = \<3,1,11,1\>$ &	\\ 
$\H(I)= \<14,2,12,2,0,10,2\>$ & $\HH(I)= \<13,3,11,4,1,12,1\>$ & $\HC(I)= \<14,3,11,3,0,11,0\>$\\
$\S(I)= \<0,2,2,10\>$ & $\SS(I)= \<1,2,0,11\>$ & $\SC(I)= \<1,1,1,11\>$ \\ \hline
\end{tabular} 
}
\caption{(a) Illustration of sample data representing a bipartite graph of network connections; (b) Definitions and sample values for alternative query sequences: $\L$ counts the number of connections for each $source$, $\H$ provides a hierarchy of range counts, and $\S$ returns an ordered degree sequence for the implied graph.}
\label{fig:example1}
\end{center}
\vspace{-0.5cm}
\end{figure*}

First, given a task---such as computing a histogram over student grades---the  analyst chooses a set of queries $\Q$ to send to the data owner.  The choice of queries will depend on the particular task, but in this work they are chosen so that constraints hold among the answers.  For example, rather than issue $(\q_A, \q_B, \q_C, \q_D, \q_F)$, the analyst would formulate the query as $(\q_t, \q_p, \q_A, \q_B, \q_C, \q_D, \q_F)$, which has consistency constraints.  The query set $\Q$ is sent to the data owner.


In the second step, the data owner answers the set of queries, using a standard differentially-private mechanism~\cite{dwork2006calibrating}, as follows.  The queries are evaluated on the private database and the true answer $\Q(\db)$ is computed.  Then random independent noise is added to each answer in the set, where the data owner scales the noise based on the sensitivity of the query set.  The set of noisy answers $\qq$ is sent to the analyst. Importantly, because this step is unchanged from~\cite{dwork2006calibrating}, it offers the same differential privacy guarantee.
%
%

The above step ensures privacy, but the set of noisy answers returned may be inconsistent.  In the third and final step, the analyst post-processes the set of noisy answers to resolve inconsistencies among them.  We propose a novel approach for resolving inconsistencies, called \emph{constrained inference}, that finds a new set of answers $\qbar$ that is the ``closest'' set to $\qq$ that also satisfies the consistency constraints.

For two histogram tasks, our main technical contributions are efficient techniques for the third step and a theoretical and empirical analysis of the accuracy of $\qbar$.  The surprising finding is that $\qbar$ can be significantly more accurate than $\qq$.

We emphasize that the constrained inference step has {\em no impact} on the differential privacy guarantee.  The analyst performs this step without access to the private data, using only the constraints and the noisy answers, $\qq$.  The noisy answers $\qq$ are the output of a differentially private mechanism; any post-processing of the answers cannot diminish this rigorous privacy guarantee.  The constraints are properties of the \emph{query}, not the database, and therefore known by the analyst \emph{a priori}.  For example, the constraint $\q_p = \q_A + \q_B + \q_C + \q_D$ is simply the definition of $\q_p$.

Intuitively, however, it would seem that if noise is added for privacy and then constrained inference reduces the noise, some privacy has been lost.  In fact, our results show that existing techniques add more noise than is strictly necessary to ensure differential privacy.  The extra noise provides no quantifiable gain in privacy but does have a significant cost in accuracy.  We show that constrained inference can be an effective strategy for boosting accuracy.

\eat{
In this paper we propose novel techniques for deriving private but accurate answers to two related histogram tasks.  Our approach is the following.  First, we choose a set of queries where constraints naturally hold among the query answers.  Second, using a standard differentially private algorithm, we derive a set of noisy answers.  If the true answer set is $x$, we write the noisy answer set $\widetilde{x}$.  The noisy answer set typically violates the constraints.  So, third, we introduce a {\em constrained inference} step that computes a new set of answers $\overline{x}$ that satisfies the constraints and is closest to $\widetilde{x}$.  The inferred answers $\overline{x}$ are differentially private and consistent, and often significantly more accurate than the output $\widetilde{x}$.  

We emphasize that the improvement in accuracy is achieved with {\em no sacrifice of privacy}. The constrained inference step is applied by the analyst after receiving the output of a differentially private algorithm.  In addition, the consistency constraints used in the inference step are derived from properties of the queries and contain no information about the private database instance.  Since the inference step uses no information from the database other than $\widetilde{x}$, it has no impact on the privacy guarantee.  Thus, all techniques proposed in this paper satisfy standard $\epsilon$-differential privacy \cite{dwork2008differential}. 
}

The increase in accuracy we achieve depends on the input database and the privacy parameters.  For instance, for some databases and levels of noise the perturbation may tend to produce answers that do not violate the constraints.  In this case the inference step would not improve accuracy.  But we show that our inference process never reduces accuracy and give conditions under which it will boost accuracy.  In practice, we find that many real datasets have data distributions for which our techniques significantly improve accuracy.

\eat{
Computing consistent answers from differentially private outputs has been discussed before \cite{barak2007privacy} for the special case of contingency tables, although accuracy is not improved.  We are able to exploit a larger set of constraints, and more general constraints, which offers a boost in accuracy.  
}


%


\eat{Post-processing the noisy output of a differentially private algorithm is in fact a common practice.  A noisy output may fail basic constraints like integrality and non-negativity.  Typically the output is simply rounded.  This can be seen as a special case of our approach, but one for which constrained inference is not hard to do and does not add much to utility.

Consistency was explicitly considered \cite{barak2007privacy} for contingency tables (which are more general than our histograms).   Similar to our approach, a standard differentially private algorithm is used to get initial output.  That output is post-processed to enforce mutual consistency.  The techniques do not increase accuracy (in fact, they decrease accuracy, although it is shown theoretically that the loss in accuracy is usually modest).}

\paragraph*{\bf Histogram tasks}

We demonstrate this technique on two specific tasks related to histograms. 
For relational schema $R(A, B, \dots)$, we choose one attribute $A$ on which histograms are built (called the {\em range} attribute).  We assume the domain of $A$, $dom$, is ordered.  

We explain these tasks using sample data that will serve as a running example throughout the paper, and is also the basis of later experiments.  The relation $R(src,dst)$, shown in Fig. \ref{fig:example1}, represents a trace of network communications between a source IP address ({\em src}) and a destination IP address ({\em dst}).  It is bipartite because it represents flows through a gateway router from internal to external addresses.

In a conventional histogram, we form disjoint intervals for the range attribute and compute counting queries for each specified range.  In our example, we use $src$ as the range attribute.  There are four source addresses present in the table.  If we ask for counts of all unit-length ranges, then the histogram is simply the sequence $\< 2, 0, 10, 2 \>$ corresponding to the (out) degrees of the source addresses $\< 000,001,010,011 \>$.

Our first histogram task is an \textbf{unattributed histogram}, in which the intervals themselves are irrelevant to the analysis and so we report only a multiset of frequencies.  For the example histogram, the multiset is $\set{0, 2, 2, 10}$.  An important instance of an unattributed histogram is the degree sequence of a graph, a crucial measure that is widely studied \cite{Newman:2003The-Structure-and-Function}.  If the tuples of $R$ represent queries submitted to a search engine, and $A$ is the search term, then an unattributed histogram shows the frequency of occurrence of all terms (but not the terms themselves), and can be used to study the distribution. 



For our second histogram task, we consider more conventional sequences of counting queries in which the intervals studied may be irregular and overlapping.  In this case, simply returning unattributed counts is insufficient.  And because we cannot predict ahead of time all the ranges of interest, our goal is to compute privately a set of statistics sufficient to support arbitrary interval counts and thus any histogram.  We call this a \textbf{universal histogram}.

Continuing the example, a universal histogram allows the analyst to count the number of packets sent from any single address (e.g., the counts from source address 010 is 10), or from any range of addresses (e.g., the total number of packets is 14, and the number of packets from a source address matching prefix $01*$ is 12). 


While a universal histogram can be used compute an unat\-tributed histogram, we distinguish between the two because we show the latter can be computed much more accurately.  

\eat{ 
\subsection*{Motivating Example}
\begin{outline}[enumerate]
\1 (Histogram) An analyst who wants to count the number of connections for each source address can ask for a sequence of counts, denoted $\L:\<C_{000},C_{001},C_{010},C_{011}\>$, where $C_{000}$ returns the number of tuples in $R$ for which $R.A = 000$.

\2 For the sample data, the true answer is $\L(I)=\<2,0,10,2\>$.  The privacy mechanism returns a noisy answer, denoted $\LL(I) = \<3,1,11,1\>$.  Each count is off by 1 in this simplified example.  The analyst can combine individual counts to get estimates of other bucket counts.  For example, using $\L(I)$, the estimate for $C_{0**}$, which is the sum over all addresses, is the sum of the components: 16.

\2 Our technique for histograms asks a different query, denoted $\H$.  It consists of counts for each address, but also a hierarchical counts: $C_{00*}$ the count of addresses in the left subtree, $C_{01*}$ the count of addresses in the right subtree, and the count of all addresses $C_{0**}$.  The true answer is $\H(I)= \<14,2,12,2,0,10,2\>$.  To satisfy differential privacy we must add more noise to each of the 7 query answers.  The noisy answer we get is $\HH(I)= \<13,3,11,4,1,12,1\>$, in which some statistics are off by 2.  But the mutual constraints that hold for $\H(I)$ are violated in $\HH(I)$.  Our inference process applies the constraints to get 
$\HC(I)= \<13,2,11,2,0,10,1\>$, which is consistent.  $\HC(I)$ is guaranteed to be more accurate than $\HH(I)$, but more importantly, it more accurate than $\LL(I)$ (for higher nodes, and often lower nodes too). 

\2 In Sec~\ref{sec:experiments} we apply this technique to a real network trace and find that accuracy is improved by XX\%.

\1 (Unattributed Histogram) If an analyst wants to study just the distribution of activity across the network, then the identities of the source addresses may be irrelevant.  For such analyses, the degree distribution is sufficient. 

\2 The conventional technique is simply to issue query $\L$, discussed above.  This gives the out-degree for each source node. 

\2 We propose a novel query, denoted $\S$, which returns the counts in sorted order.  More precisely, $\S:\<Rank_1, Rank_2, Rank_3, Rank_4\>$ where $Rank_i$ returns the count of the source address whose count has rank $i$.  Structuring the query this way has a significant impact.  First, the noise added to privately answer this query is {\em not} larger than $\L$.  Second, $\S$ has mutual constraints that can be used in the inference step to improve accuracy.

\2 In Sec~\ref{sec:experiments} we apply this technique to a real network trace and find that accuracy is improved by YY\%.  We believe this is the most accurate differentially private technique for the estimation of degree distributions, and could have important applications.  
\end{outline}
} 

\paragraph*{\bf Contributions}
For both unattributed and universal histograms, we propose a strategy for boosting the accuracy of existing differentially private algorithms. 
For each task, (1) we show that there is an efficiently-computable, closed-form expression for the {\em consistent} query answer closest to a private randomized output; (2) we prove bounds on the error of the inferred output, showing under what conditions inference boosts accuracy; \eat{, and in some cases, that it provides an optimal estimate given the input;} (3) we demonstrate significant improvements in accuracy through experiments on real data sets.  \eat{Degree sequence estimation}Unattributed histograms are extremely accurate, with error at least an order of magnitude lower than existing techniques.  Our approach to universal histograms can reduce error for larger ranges by 45-98\%, and improves on all ranges in some cases.

\eat{Our results show that common differential privacy approaches can add more noise than is strictly required by the privacy condition.  Using constraints and the proposed inference process is an important method for avoiding this.}

\section{Background} \label{sec:background}

\hyphenation{un-attr-ibuted} 

In this section, we introduce the concept of query sequences and how they can be used to support histograms.  Then we review differential privacy and show how queries can be answered under differential privacy.  Finally, we formalize our constrained inference process.

All of the tasks considered in this paper are formulated as {\em query sequences} where each element of the sequence is a simple count query on a range.  We write intervals as $[x,y]$ for $x,y \in dom$, and abbreviate interval $[x,x]$ as $[x]$.  A counting query on range attribute $A$ is:\\

$\cnt([x,y])$ = {\sf Select count(*) From R Where $x\leq$ R.A $\leq y$} \\

\eat{Our notational conventions are summarized in Table \ref{tbl:notation}.}
\eat{We refer to query sequences using bold letters ($\Q,\L,\H,\S$).} 

We use $\Q$ to denote a generic query sequence (please see Appendix~\ref{sec:conventions} for an overview of notational conventions).  When $\Q$ is evaluated on a database instance $I$, the output, $\Q(I)$, includes one answer to each counting query, so $\Q(I)$ is a vector of non-negative integers.  The $i^{th}$ query in $\Q$ is $\Q[i]$.

We consider the common case of a histogram over unit-length ranges.  The conventional strategy is to simply compute counts for all unit-length ranges.  This query sequence is denoted $\L$:
$$\L = \< \: \cnt([x_1]), \dots \cnt([x_n])\:\>, x_i \in dom$$

\begin{example}
	Using the example in Fig~\ref{fig:example1}, we assume the domain of $src$ contains just the 4 addresses shown. Query $\L$ is 
	$\< \: \cnt([000]), \cnt([001]),\cnt([010]), \cnt([011])\:\>$
	and $\L(I)=\<2,0,10,2\>$.
\end{example}

\eat{  
\subsection{Preliminaries}
We consider a private database $I$, which is an instance of relational schema $R(A, B, \dots)$.  We assume attribute $A$ is ordered with $|dom(A)|=n$. We consider counting queries on attribute $A$, denoted $\cnt([x,y])$, which return, for $x,y \in dom$, a non-negative integer counting all tuples $t$ for which $x \leq t.A \leq y$.  We use $\cnt([x])$ to abbreviate $\cnt([x,x])$.

}

\subsection{Differential Privacy} \label{sec:sub:dp}


Informally, an algorithm is differentially private if it is insensitive to small changes in the input. 
Formally, for any input database $\db$, let $\nbrs(\db)$ denote the set of neighboring databases, each differing from $\db$ by at most one record; i.e., if $\db' \in \nbrs(\db)$, then $|(\db - \db') \cup (\db' - \db)| = 1$.
%
%
%
	\begin{definition}[$\epsilon$-differential privacy] Algorithm $A$ is $\epsilon$-differentially private if for all instances $I$, any $I' \in \nbrs(I)$, and any subset of outputs $S \subseteq Range(A)$, the following holds:
\[
Pr[ A(I) \in S] \leq \exp(\epsilon) \times Pr[ A(I') \in S]
\]		
where the probability is taken over the randomness of the $A$.
	\end{definition}
Differential privacy has been defined inconsistently in the literature.  The original concept, called $\epsilon$-indistinguishability~\cite{dwork2006calibrating}, defines neighboring databases using hamming distance rather than symmetric difference (i.e., $\db'$ is obtained from $\db$ by \emph{replacing} a tuple rather than adding/removing a tuple).  The choice of definition affects the calculation of query sensitivity.  We use the above definition (from Dwork~\cite{dwork2008differential}) but observe that our results also hold under indistinguishability, due to the fact that $\epsilon$-differential privacy (as defined above) implies $2\epsilon$-indistinguishability.



To answer queries under differential privacy, we use the Laplace mechanism~\cite{dwork2006calibrating}, which achieves differential privacy by adding noise to query answers, where the noise is sampled from the Laplace distribution.  The magnitude of the noise depends on the query's \emph{sensitivity}, defined as follows (adapting the definition to the query sequences considered in this paper).

\begin{definition}[Sensitivity]  Let $\Q$ be a sequence of counting queries.  The sensitivity of $\Q$, denoted $S_{\Q}$, is
\[
\sens{\Q} = \max_{\db,\db' \in \nbrs(\db)} \Lone{\Q(\db)}{\Q(\db')} 
\]
\end{definition}
Throughout the paper, we use $\Ldist{\mathbf{X}}{\mathbf{Y}}{p}$ to denote the $L_p$ distance between vectors $\mathbf{X}$ and $\mathbf{Y}$.

\begin{example}
	The sensitivity of query $\L$ is 1.  Database $\db'$ can be obtained from $\db$ by adding or removing a single record, which affects exactly one of the counts in $\L$ by exactly 1.
\end{example}

Given query $\Q$, the Laplace mechanism first computes the query answer $\Q(\db)$ and then adds random noise independently to each answer.  The noise is drawn from a zero-mean Laplace distribution with scale $\sigma$.  As the following proposition shows, differential privacy is achieved if the Laplace noise is scaled appropriately to the sensitivity of $\Q$ and the privacy parameter $\epsilon$.  

\begin{proposition}[Laplace mechanism~\cite{dwork2006calibrating}]
	\label{prop:laplace}
	Let $\Q$ be a query sequence of length $d$, and let $\<\Lap(\sigma)\>^d$ denote a $d$-length vector of i.i.d. samples from a Laplace with scale $\sigma$.  
	The randomized algorithm $\QQ$ that takes as input database $\db$ and outputs the following vector is $\epsilon$-differentially private: $$\QQ(I) = \Q(I) + \<\Lap( \sens{\Q}/\epsilon )\>^d$$
\end{proposition}

We apply this technique to the query $\L$.  Since, $\L$ has sensitivity 1, the following algorithm is $\epsilon$-differentially private:
$$\LL(I) = \L(I) + \<\Lap(1/\epsilon)\>^n$$

We rely on Proposition~\ref{prop:laplace} to ensure privacy for the query sequences we propose in this paper.  We emphasize that the proposition holds for \emph{any} query sequence $\Q$, regardless of correlations or constraints among the queries in $\Q$.  Such dependencies are accounted for in the calculation of sensitivity.  (For example, consider the correlated sequence $\Q$ that consists of the \emph{same} query repeated $k$ times, then the sensitivity of $\Q$ is $k$ times the sensitivity of the query.)

We present the case where the analyst issues a single query sequence $\Q$.  To support multiple query sequences, the protocol that computes a $\epsilon_i$-differentially private response to the $i^{th}$ sequence is $(\sum \epsilon_i)$-differentially private.



To analyze the accuracy of the randomized query sequences proposed in this paper we quantify their error.  $\QQ$ can be considered an estimator for the true value $\Q(I)$.  We use the common Mean Squared Error as a measure of accuracy. 
\begin{definition}[Error] \label{def:error} 
For a randomized query sequence $\QQ$ whose input is $\Q(I)$, 
the $\error(\QQ)$ is $\sum_i \E (\QQ[i] - \Q[i])^2$ 
Here $\E$ is the expectation taken over the possible randomness in generating $\QQ$.
\end{definition}
For example, $\error(\LL) = \sum_i \E (\LL[i] - \L[i])^2$ which simplifies to: $n \, \E [ Lap(1/\epsilon)^2 ] = n \, Var( Lap(1/\epsilon) ) = 2n/\epsilon^2$.

\subsection{Constrained Inference}

\eat{The queries $\S$ and $\H$ for unattributed and universal histograms were described in Sec~\ref{sec:sub:construction} along with their constraints $\cons_{\S}$ and $\cons_{\H}$. The differentially private extensions $\SS$ and $\HH$ are described in Sec~\ref{sec:unattributed} and \ref{sec:hierarchical}, respectively.}

While $\LL$ can be used to support unattributed and universal histograms under differential privacy, the main contribution of this paper is the development of more accurate query strategies based on the idea of constrained inference.  The specific strategies are described in the next sections.  Here, we formulate the constrained inference problem.

Given a query sequence $\Q$, let $\cons_{\Q}$ denote a set of constraints which must hold among the (true) answers. 
\eat{Recalling the example from the introduction, 
$\cons_{\Q}$ would consist of the constraints $\q_t = \q_p + \q_F$ and $\q_p = \q_A + \q_B + \q_C + \q_D$.  }
 The constrained inference process takes the randomized output of the query, denoted $\qq=\QQ(I)$, and finds the sequence of query answers $\qbar$ that is ``closest'' to $\qq$ and also satisfies the constraints of $\cons_{\Q}$. 
Here closest is determined by $L_2$ distance, and the result is the {\em minimum $L_2$ solution}: 
\begin{definition}[Minimum $L_2$ solution] \label{def:minL2} 
Let $\Q$ be a query sequence with constraints $\cons_{\Q}$.  Given a noisy query sequence $\qq=\QQ(I)$, a minimum $L_2$ solution, denoted 
$\qbar$, is a vector $\qbar$ that satisfies the constraints $\cons_{\Q}$ and at the same time minimizes $\Ltwo{\qq}{\qbar}$.  
\end{definition}


We use $\QC$ to denote the two step randomized process in which the data owner first computes $\qq = \QQ(I)$ and then computes the minimum $L_2$ solution from $\qq$ and $\cons_{\Q}$.  (Alternatively, the data owner can release $\qq$ and the latter step can be done by the analyst.)  Importantly, the inference step has no impact on privacy, as stated below.  (Proofs appear in the Appendix.)
\begin{restatable}{proposition}{transformationInvariance}
	\label{prop:inference_privacy}
	If $\QQ$ satisfies $\epsilon$-differential privacy, then $\QC$ satisfies $\epsilon$-differential privacy.
\end{restatable}

\cut{
The query sequences that we propose in Sec~\ref{sec:unattributed} \& \ref{sec:hierarchical} have unique minimum $L_2$ solutions.  
Uniqueness is one of the main reasons we choose to formulate the inference process using $L_2$ distance; with $L_1$ distance, the solution may not be unique or as accurate.  In addition, we shall show that the minimal $L_2$ solution has maximum utility in many cases.

Since the query sequences considered in the paper consist of counting queries, the counts of the true sequence are always integral and non-negative.  However, the counts of the noisy sequence may not be.  While incorporating integrality and non-negativity constraints may further boost accuracy, to simplify theoretical development, we do not include them in $\cons_{\Q}$.  We do consider them in the experiments of  Sec~\ref{sec:experiments}.
}

\eat{
\vr{Is the paragraph below needed? I am not sure about the fact that differential privacy can be adjusted to work with L2 sensitivity} The attentive reader will note that the definition of sensitivity uses the $L_1$ distance.  Although differential privacy can be re-defined using $L_2$ distance, it is not really a problem to use a different distance metric for the post-processing, so we choose to stay with the classical formulation of differential privacy.
}

\eat{
\paragraph*{Review of notational conventions} Bold capital letters are used to refer to deterministic query sequences ($\L,\H,\S$).  When evaluated on an instance, ($\L(I),\H(I),\S(I)$) they return the true answer to the query sequence (a vector of numbers). With tilde, ($\LL,\HH,\SS$) they refer to randomized query sequences, which can be viewed as estimators for the true answers.  The noisy output, also a vector of numbers, is denoted $\lsmall=\LL(I), \hh=\HH(I), \ss=\SS(I)$.

 they refer to deterministic query sequences, whose components are counting queries. 
 
   With bar, ($\HC,\SC$) they refer to randomized queries which result from applying post-processing.  So we have $\HC$ defined to be $\minTwo(\HH(I),\cons_{\H})$.  Since the $\minTwo$ is unique, I think it makes sense to consider this a randomized query, but it may be confusing. 

Any such query can be evaluated on an instance $I$ and will return an output sequence, that is, a vector of numbers. So we can write $\S(I)$ for the true answer, $\SS(I)$ for the private answer, and $\minTwo(\SS(I),\cons_{\S})$ for the inferred answer (or perhaps $\SC(I)$, but I've avoided this).  I've recently introduced small letters for output sequences.  I use $\ss=\SS(I)$ and $\sbar=\SC(I)$ in Sec 3.

Both query sequences and output sequences are vectors, so we can use vector notation.  $\Q[i]$ is a counting query, $\QQ[i]$ is a counting query with laplace noise, etc. And $\ss[i]$ is a single noisy count, $\sbar[i]$ is an inferred count.  This avoids having to write $\SS(I)[i]$.  
}

\section{Unattributed histograms}
\label{sec:unattributed}

To support unattributed histograms, one could use the query sequence $\L$.  However, it contains ``extra'' information---the attribution of each count to a particular range---which is irrelevant for an \emph{unattributed} histogram.  Since the association between $\L[i]$ and $i$ is not required, any permutation of the unit-length counts is a correct response for the unattributed histogram.  We formulate an alternative query that asks for the counts of $\L$ in rank order.  As we will show, ordering does not increase sensitivity, but it does introduce inequality constraints that can be exploited by inference.

\def\answers{\mathcal{U}}
\def\subAnswers{\mathcal{V}}
Formally, let $a_i$ refer to the answer to $\L[i]$ and let $\answers = \{ a_1, \dots, a_n \}$ be the multiset of answers to queries in $\Q$.  We write $rank_i(\answers)$ to refer to the $i^{th}$ smallest answer in $\answers$.  
Then the query $\S$ is defined as 
\[
\S = \< rank_1(\answers), \dots, rank_n(\answers) \>
\]


\begin{example}
	In the example in Fig~\ref{fig:example1}, we have $\L(\db)=\<2,0,10,2\>$ while $\S(\db)= \<0,2,2,10\>$.  Thus, the answer $\S(\db)$ contains the same counts as $\L(\db)$ but in \emph{sorted} order.
\end{example}
To answer $\S$ under differential privacy, we must determine its sensitivity.  %
\begin{restatable}{proposition}{sensitivityS}
	\label{prop:sensitivity_s}
	The sensitivity of $\S$ is 1.
\end{restatable} 
By Propositions~\ref{prop:laplace} and \ref{prop:sensitivity_s}, the following algorithm is $\epsilon$-differentially private:
\begin{eqnarray*}
	\SS(\db) &=& \S(\db) + \<\Lap(1/\epsilon)\>^n
\end{eqnarray*}
Since the same magnitude of noise is added to $\S$ as to $\L$, the accuracy of $\SS$ and $\LL$ is the same.  However, $\S$ implies a powerful set of constraints.  Notice that the ordering occurs \emph{before} noise is added.  Thus, the analyst knows that the returned counts are ordered according to the true rank order.  If the returned answer contains out-of-order counts, this must be caused by the addition of random noise, and they are inconsistent.  Let $\cons_{\S}$ denote the set of inequalities $\S[i] \leq \S[i+1]$ for $1 \leq i < n$.  We show next how to exploit these constraints to boost accuracy.

\subsection{Constrained Inference: Computing $\SC$}
\label{sec:inference_s}

As outlined in the introduction, the analyst sends query $\S$ to the data owner and receives a noisy answer $\ss = \SS(I)$, the output of the differentially private algorithm $\SS$ evaluated on the private database $\db$.  
We now describe a technique for post-processing $\ss$ to find an answer that is consistent with the ordering constraints.

Formally, given $\ss$, the objective is to find an $\sbar$ that minimizes $\Ltwo{\ss}{\sbar}$ subject to the constraints $\sbar[i] \leq \sbar[i+1]$ for $1 \leq i < n$.  The solution has a surprisingly elegant closed-form.  Let $\ss[i,j]$ be the subsequence of $j-i+1$ elements: $\<\ss[i]$, $\ss[i+1]$, $\ldots, \ss[j]\>$.  Let $\mm[i,j]$ be the mean of these elements, i.e. $\mm[i,j] = \sum_{k=i}^j \ss[k] / (j-i+1)$.

\begin{restatable}{theorem}{minLTwoOC}
	\label{thm:minL2OC}
	Denote $L_k = \min_{j \in [k,n]} \max_{i \in [1,j]} \mm[i,j]$ and $U_k =  \max_{i\in [1,k]} \min_{j\in [i,n]} \mm[i,j]$. The minimum $L_2$ solution 
	$\sbar$, is unique and given by: $\sbar[k] = L_k = U_k$.
\end{restatable}

Since we first stated this result in a technical report~\cite{hay2009boosting0}, we have learned that this problem is an instance of isotonic regression (i.e., least squares regression under ordering constraints on the estimands).  The statistics literature 
gives several characterizations of the solution, including the above min-max formulas (cf. Barlow et al.~\cite{barlow1972the-isotonic}), as well as linear time algorithms for computing it (cf. Barlow et al.~\cite{barlow1972statistical}).

\begin{example}
We give three examples of $\ss$ and its closest ordered sequence $\sbar$.  
	First, suppose $\ss=\<9, 10, 14\>$.  Since $\ss$ is already ordered, $\sbar = \ss$.  Second, $\ss=\<9, 14, 10\>$, where the last two elements are out of order.  The closest ordered sequence is $\sbar=\<9, 12, 12\>$.  Finally, let $\ss=\<14, 9, 10, 15\>$.  The sequence is in order except for $\ss[1]$.  While changing the first element from 14 to 9 would make it ordered, its distance from $\ss$ would be $(14-9)^2 = 25$.  In contrast, $\sbar = \<11, 11, 11, 15\>$ and $\Ltwo{\ss}{\sbar} = 14$.
\end{example}

\subsection{Utility Analysis: the Accuracy of $\SC$}

Prior work in isotonic regression has shown inference cannot hurt, i.e., the accuracy of $\SC$ is no lower than $\SS$~\cite{hwang1994confidence}.  However, we are not aware of any results that give conditions for which $\SC$ is more accurate than $\SS$.  Before presenting a theoretical statement of such conditions, we first give an illustrative example.
\begin{figure}[ht]
\begin{center}
\includegraphics[width=3.2in]{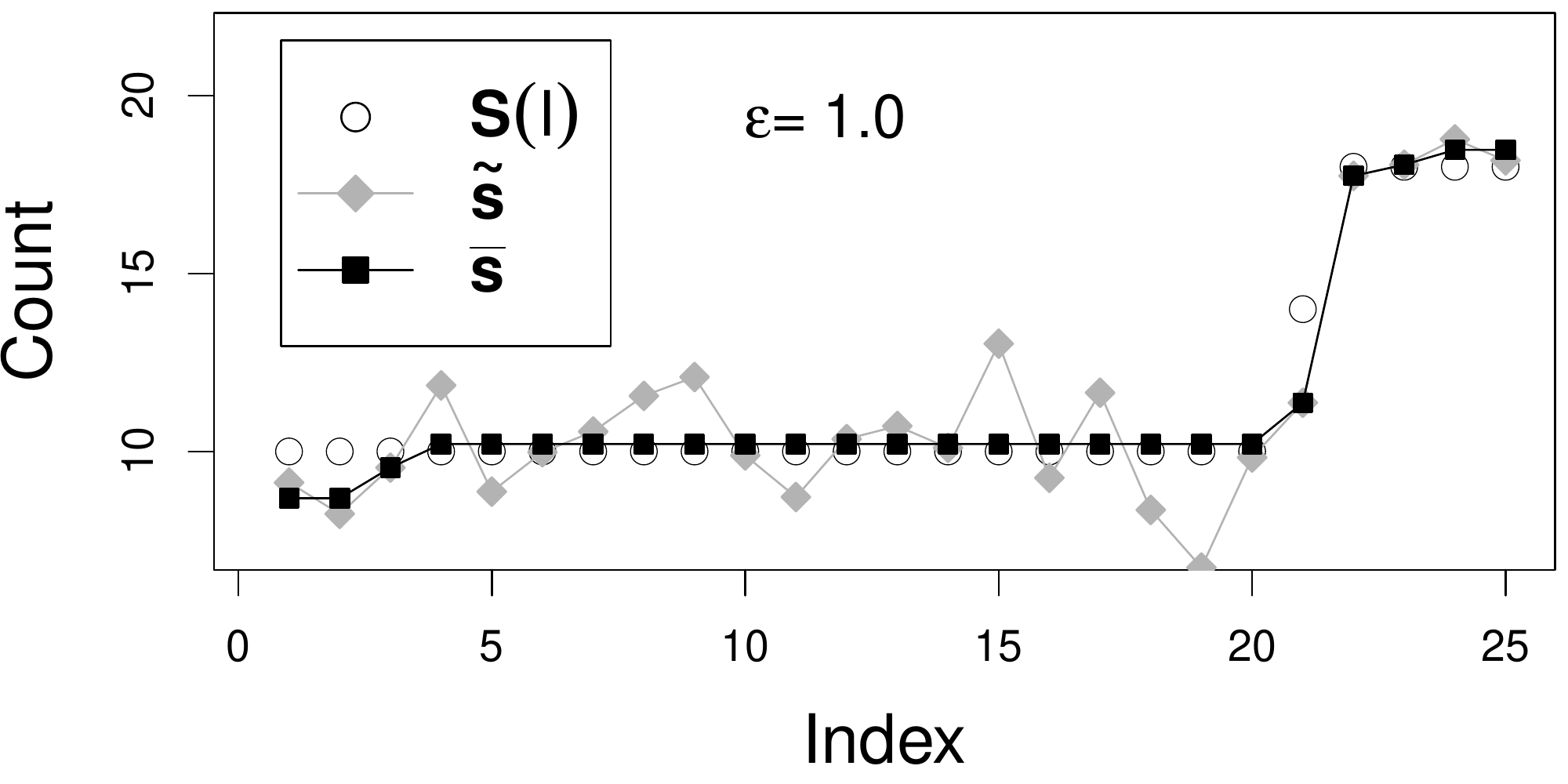}
\caption{
\label{fig:sequence_error_toy} Example of how $\sbar$ reduces the error of $\ss$.
}
\end{center}
\vspace{-0.7cm}
\end{figure}
\begin{example}
Figure~\ref{fig:sequence_error_toy} shows a sequence $\S(I)$ along with a sampled $\ss$ and inferred $\sbar$.  While the values in $\ss$ deviate  considerably from $\S(I)$, $\sbar$ lies very close to the true answer.  In particular, for subsequence $[1,20]$, the true sequence $\S(I)$ is uniform and the constrained inference process effectively averages out the noise of $\ss$.  At the twenty-first position, which is a unique count in $\S(I)$, and constrained inference does not refine the noisy answer, i.e., $\sbar[21] = \ss[21]$.
\end{example}

%

Fig~\ref{fig:sequence_error_toy} suggests that $\error(\SC)$ will be low for sequences in which many counts are the same (Fig~\ref{fig:sequence_error} in Appendix~\ref{sec:app_unattributed} gives another intuitive view of the error reduction).  
The following theorem quantifies the accuracy of $\SC$ precisely.  Let $n$ and $d$ denote the number of values and the number of distinct values in $\S(I)$ respectively. Let $n_1,n_2,\ldots,n_d$ be the number of times each of the $d$ distinct values occur in $\S(I)$~(thus $\sum_i n_i = n$).

\begin{restatable}{theorem}{thmUnattributed}
	\label{theorem:unattributed}
	There exist constants $c_1$ and $c_2$ independent of $n$ and $d$ such that 
\begin{eqnarray*} 
\error(\SC) \leq  \sum_{i=1}^d \frac{c_1\log^3{n_i}+c_2}{\epsilon^2}
\end{eqnarray*}
Thus $\error(\SC) = O(d \log^3{n}/\epsilon^2)$ whereas $\error(\SS) = \Theta(n/\epsilon^2)$.
\end{restatable}

The above theorem shows that constrained inference can boost accuracy, and the improvement depends on properties of the input $\S(\db)$.  In particular, if the number of distinct elements $d$ is $1$, then $\error(\SC) = O(\log^3{n}/\epsilon^2)$, while $\error(\SS) = \Theta(n/\epsilon^2)$. On the other hand, if $d=n$, then $\error(\SC)=O(n/\epsilon^2)$ and thus both $\error(\SC)$ and $\error(\SS)$ scale linearly in $n$.  For many practical applications, $d \ll n$, which makes $\error(\SC)$ significantly lower than $\error(\SS)$.  In Sec.~\ref{sec:experiments}, experiments on real data demonstrate that the error of $\SC$ can be orders of magnitude lower than that of $\SS$.


\eat{

The bound for $\error(\SC)$ as shown in statement $(ii)$ of the above theorem is perhaps weaker than the actual bound by a factor of $\log{n}$, as can be seen from the case when $d = n$ in which case $(i)$ implies that $\error(\SC) = O(n/\epsilon^2)$. However, for the general case of $d$, $(ii)$ is the best bound we can obtain, which could be a limitation of the analysis in our proof rather than a property of $\error(\SC)$. The above  $\log{n}$.\\
}


\section{Universal histograms} 
\label{sec:hierarchical}

While the query sequence $\L$ is the conventional strategy for computing a universal histogram, this strategy has limited utility under differential privacy.  While accurate for small ranges, the noise in the unit-length counts accumulates under summation, so for larger ranges, the estimates can easily become useless.  

We propose an alternative query sequence that, in addition to asking for unit-length intervals, asks for interval counts of larger granularity.  To ensure privacy, slightly more noise must be added to the counts.  However, this strategy has the property that any range query can be answered via a linear combination of only a small number of noisy counts, and this makes it much more accurate for larger ranges.

Our alternative query sequence, denoted $\H$, consists of a sequence of hierarchical intervals.  
Conceptually, these intervals are arranged in a tree $T$.  
Each node $v \in T$ corresponds to an interval, and each node has $k$ children, corresponding to $k$ equally sized subintervals.  The root of the tree is the interval $[x_1, x_n]$, which is recursively divided into subintervals until, at leaves of the tree, the intervals are unit-length, $[x_1],[x_2],\dots, [x_n]$.  
For notational convenience, we define the height of the tree $\hght$ as the number of \emph{nodes}, rather than edges, along the path from a leaf to the root.  Thus, $\hght = \log_k n + 1$.
To transform the tree into a sequence, we arrange the interval counts in the order given by a breadth-first traversal of the tree.

\begin{figure}[h]
\begin{center}
\includegraphics[width=1.8in]{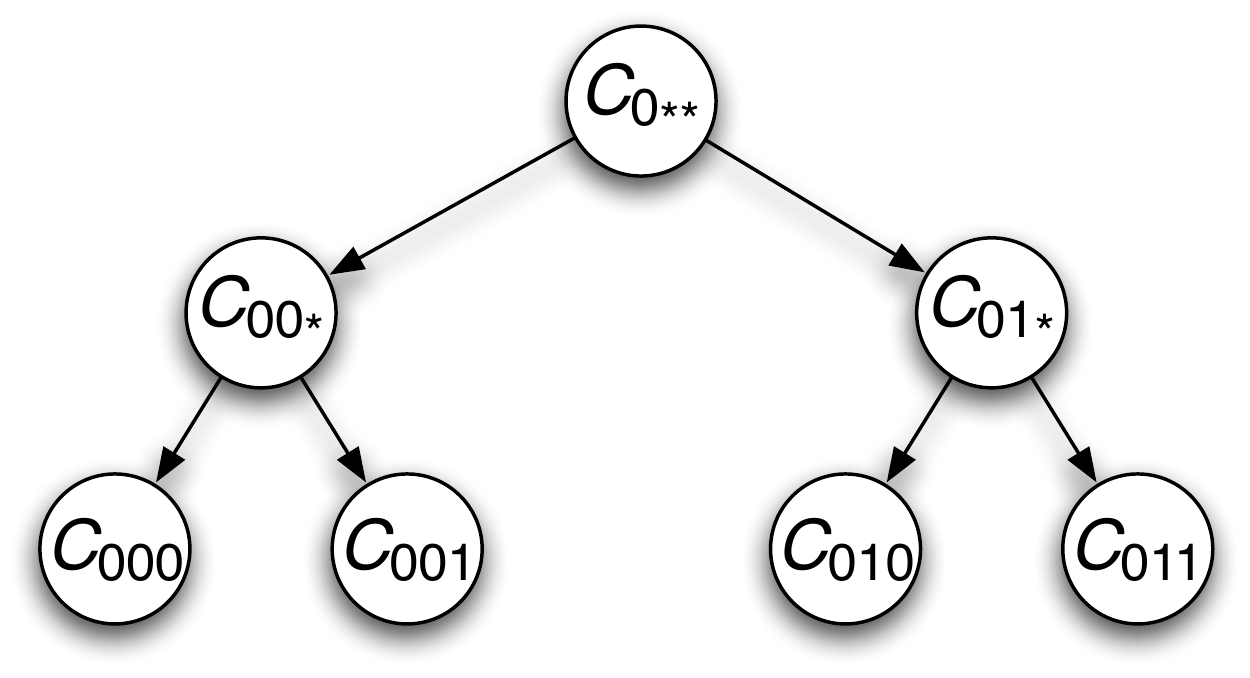}
\caption{
\label{fig:h_query} The tree $T$ associated with query $\H$ for the example in Fig.~\ref{fig:example1} for $k=2$.
}
\end{center}
\vspace{-0.7cm}
\end{figure}

\begin{example} 
	\label{ex:tree}
	Continuing from the example in Fig~\ref{fig:example1}, we describe $\H$ for the $src$ domain.  The intervals are arranged into a binary $(k=2)$ tree, as shown in Fig~\ref{fig:h_query}.  The root is associated with the interval $[0**]$, which is evenly subdivided among  its children. The unit-length intervals at the leaves are $[000], [001], [010],[011]$.  The height of the tree is $\hght=3$.
	
	The intervals of the tree are arranged into the query sequence  
	$\H = \<C_{0**}, C_{00*}, C_{01*},C_{000},C_{001},C_{010},C_{011}\>$.  Evaluated on instance $I$ from Fig.~\ref{fig:example1}, the answer is  
	$\H(I)= \<14,2,12,2,0,10,2\>$.
\end{example}
To answer $\H$ under differential privacy, we must determine its sensitivity.  As the following proposition shows, $\H$ has a larger sensitivity than $\L$.
\begin{restatable}{proposition}{sensitivityH}
	\label{prop:sensitivity_h}
	The sensitivity of $\H$ is $\hght$.
\end{restatable} 

By Propositions~\ref{prop:laplace} and \ref{prop:sensitivity_h}, the following algorithm is $\epsilon$-differentially private:
$$\HH(\db) = \H(\db) + \<\Lap(\hght/\epsilon)\>^m$$	
where $m$ is the length of sequence $\H$, equal to the number of counts in the tree.

To answer a range query using $\HH$, a natural strategy is to sum the fewest number of sub-intervals such that their union equals the desired range.  However, one challenge with this approach is inconsistency:  in the corresponding tree of noisy answers, there may be a parent count that does not equal to the sum of its children.  This can be problematic: for example, an analyst might ask one interval query and then ask for a sub-interval and receive a \emph{larger} count.

We next look at how to use the arithmetic constraints between parent and child counts (denoted $\cons_{\H}$) to derive a consistent, and more accurate, estimate $\HC$.

\subsection{Constrained Inference: Computing $\HC$}
\label{sec:inference_h}

The analyst receives $\hh = \HH(I)$, the noisy output from the differentially private algorithm $\HH$.  
%
We now consider the problem of finding the minimum $L_2$ solution: to find the $\hbar$ that minimizes $\Ltwo{\hh}{\hbar}$ and also satisfies the consistency constraints $\cons_{\H}$.

This problem can be viewed as an instance of linear regression.  The unknowns are the true counts of the unit-length intervals.  Each answer in $\hh$ is a fixed linear combination of the unknowns, plus random noise.  Finding $\hbar$ is equivalent to finding an estimate for the unit-length intervals.  In fact, $\hbar$ is the familiar least squares solution.

Although the least squares solution can be computed via linear algebra, the hierarchical structure of this problem instance allows us to derive an intuitive closed form solution that is also more efficient to compute, requiring only linear time.  Let $T$ be the tree corresponding to $\hh$; abusing notation, for node $v \in T$, we write $\hh[v]$ to refer to the interval associated with $v$.

%

\def\vht{{l}}

First, we define a possibly inconsistent estimate $z[v]$ for each node $v \in T$.  The consistent estimate $\hbar[v]$ is then described in terms of the $z[v]$ estimates.  $z[v]$ is defined recursively from the leaves to the root.  Let $\vht$ denote the height of node $v$ and $succ(v)$ denote the set of $v$'s children.  
\begin{eqnarray*}
	z[v] & = &
	\left\{
	\begin{array}{l}
	\hh[v],~\mbox{if v is a leaf node} \\
	\frac{k^{\vht} - k^{\vht-1}}{ k^{\vht} - 1 } 
	\hh[v] +
	\frac{k^{\vht-1} - 1}{ k^{\vht} - 1 } \sum_{u \in succ(v)} z[u]
	,~\mbox{o.w.}
	\end{array}
	\right.
\end{eqnarray*}
The intuition behind $z[v]$ is that it is a weighted average of two estimates for the count at $v$; in fact, the weights are inversely proportional to the variance of the estimates.

The consistent estimate $\hbar$ is defined recursively from the root to the leaves.  At the root $r$, $\hbar[r]$ is simply $z[r]$.  As we descend the tree, if at some node $u$, we have $\hbar[u] \neq \sum_{w \in succ(u)} z[w]$, then we adjust the values of each descendant by dividing the difference $\hbar[u] - \sum_{w \in succ(u)} z[w]$ equally among the $k$ descendants.  The following theorem states that this is the minimum $L_2$ solution.

\begin{restatable}{theorem}{minLTwoHC}
	\label{minL2HC}
Given the noisy sequence $\hh=\HH(I)$, the unique minimum $L_2$ solution, 
$\hbar$, is given by the following recurrence relation. Let $u$ be $v$'s parent:
\begin{eqnarray*}
\hbar[v] & = &
\left\{
\begin{array}{l}
z[v],~~\mbox{if v is the root}\\
z[v] + \frac{1}{k} (\hbar[u] - \sum_{w \in succ(u)} z[w]),~~\mbox{o.w.}
\end{array}
\right.
\end{eqnarray*}
\end{restatable}

%

Theorem~\ref{minL2HC} shows that the overhead of computing $\HC$ is minimal, requiring only two linear scans of the tree: a bottom up scan to compute $z$ and then a top down scan to compute the solution $\hbar$ given $z$.

\subsection{Utility Analysis: the Accuracy of $\HC$}
\label{sec:hiearchical_utility}

We measure utility as accuracy of range queries, and we compare three strategies: $\LL$, $\HH$, and $\HC$.  We start by comparing $\LL$ and $\HH$.

Given range query $q = \cnt([x,y])$, we derive an estimate for the answer as follows.  For $\LL$, the estimate is simply the sum of the noisy unit-length intervals in the range: $\LL_q = \sum_{i=x}^y \LL[i]$.  The error of each count is $2/\epsilon^2$, and so the error for the range is $\error(\LL_q) = O({(y-x)}/{\epsilon^2})$.

For $\HH$, we choose the natural strategy of summing the fewest sub-intervals of $\HH$.  Let ${r}_1, \dots, {r}_t$ be the roots of disjoint subtrees of $T$ such that the union of their ranges equals $[x,y]$.  Then $\HH_q$ is defined as $\HH_q = \sum_{i=1}^{t} \HH[r_i]$.  Each noisy count has error equal to $2\hght^2/\epsilon^2$ (equal to the variance of the added noise) and the number of subtrees is at most $2\hght$ (at most two per level of the tree), thus $\error(\HH_q) = O({\hght^3}/{\epsilon^2})$.

There is clearly a tradeoff between these two strategies.  While $\LL$ is accurate for small ranges, error grows linearly with the size of the range.  In contrast,  the error of $\HH$ is poly-logarithmic in the size of the domain (recall that $\hght = \Theta(\log n)$).  Thus, while $\HH$ is less accurate for small ranges, it is much more accurate for large ranges.  If the goal of a universal histogram is to bound worst-case or total error for all range queries, then $\HH$ is the preferred strategy.  

We now compare $\HH$ to $\HC$.  Since $\HC$ is consistent, range queries can be easily computed by summing the unit-length counts.  In addition to being consistent, it is also more accurate.  In fact, it is in some sense optimal: among the class of strategies that (a) produce unbiased estimates for range queries and (b) derive the estimate from linear combinations of the counts in $\hh$, there is no strategy with lower mean squared error than $\HC$.

\begin{restatable}{theorem}{utilityH}
	\label{theorem:hierarchical:optimal} 
$(i)$ $\HC$ is a linear unbiased estimator, $(ii)$ $\error(\HC_q) \leq \error(E_q)$ for all $q$ and for all linear unbiased estimators $E$, $(iii)$ $\error(\HC_q) = O(\hght^3/\epsilon^2)$ for all $q$, and $(iv)$ there exists a query $q$ s.t. $\error(\HC_q) \leq \frac{3}{2(\hght-1)(k-1)-k}\error(\HH_q)$.
\end{restatable}
Part (iv) of the theorem shows that $\HC$ can more accurate than $\HH$ on some range queries.  For example, in a height $16$ binary tree---the kind of tree used in the experiments---there is a query $q$ where $\HC_{q}$ is more accurate than $\HH_{q}$ by a factor of $\frac{2(\hght-1)(k-1)-k}{3}=9.33$.  

Furthermore, the fact that $\HC$ is consistent can lead to additional advantages when the domain is sparse.  We propose a simple extension to $\HC$: after computing $\hbar$, if there is a subtree rooted at $v$ such that $\hbar[v] \leq 0$, we simply set the count at $v$ and all children of $v$ to be zero.  This is a heuristic strategy; incorporating non-negativity constraints into inference is left for future work.  Nevertheless, we show in experiments, that this small change can greatly reduce error in sparse regions and can lead to $\HC$ being more accurate than $\LL$ even at small ranges.

\def\trace{\texttt{Net\-Trace}}
\def\socnet{\texttt{Social Network}}
\def\search{\texttt{Search Logs}}

\section{Experiments}
\label{sec:experiments}

We evaluate our techniques on three real datasets (explained in detail in Appendix~\ref{sec:experiments_appendix}):  \trace{} is derived from an IP-level network trace collected at a major university; \socnet{} is a graph derived from friendship relations in an online social network site; \search{} is a dataset of search query logs over time from Jan. 1, 2004 to the present.  Source code for the algorithms is available at the first author's website.

\subsection{Unattributed Histograms} 
\label{expt:unattributed_histograms}

\begin{figure}[!t]
\begin{center}
\includegraphics[width=3.25in]{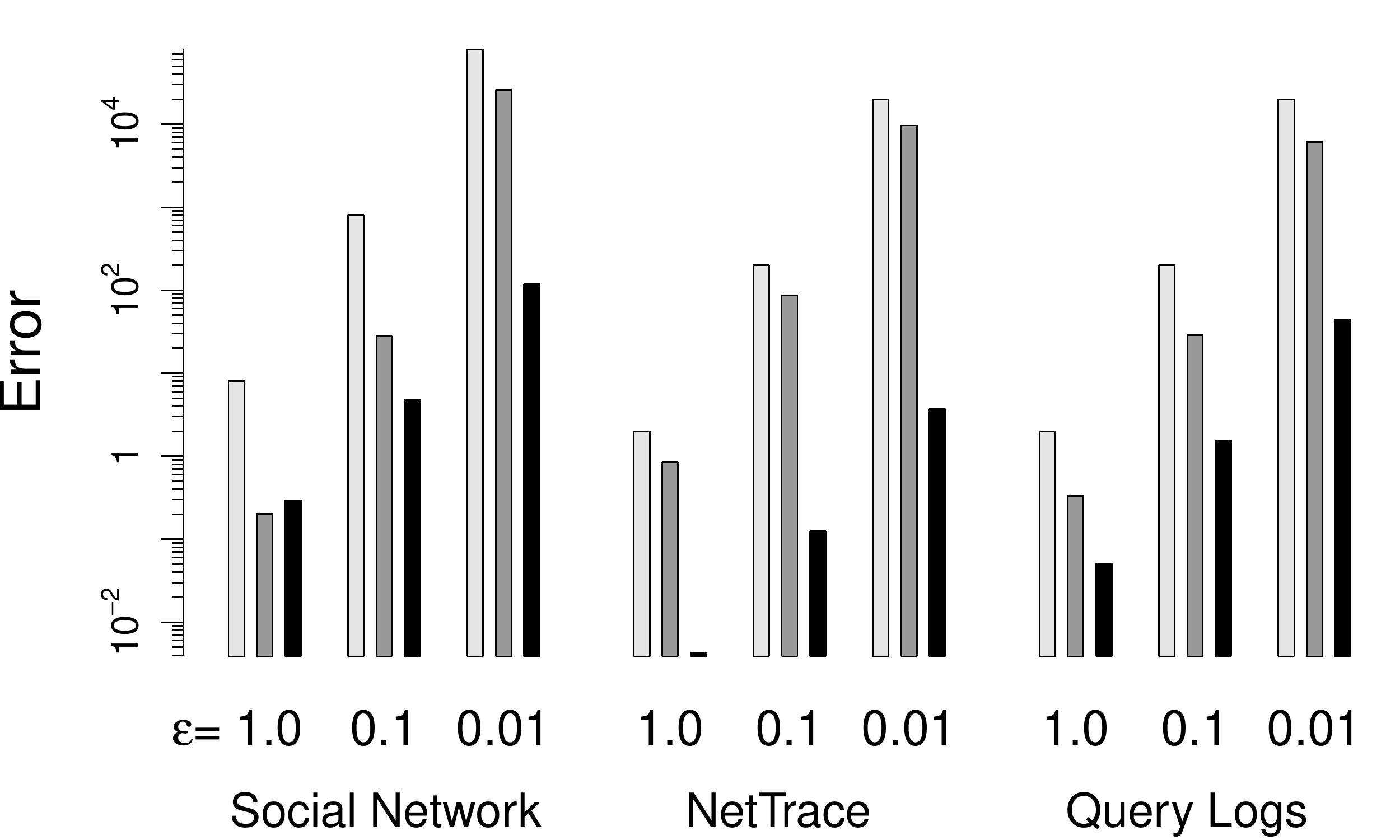}
\caption{
\label{fig:error_log} Error across varying datasets and $\epsilon$.  Each triplet of bars represents the three estimators: $\SS$ (light gray), $\SR$ (gray), and $\SC$ (black).
}
\end{center}
\vspace{-0.7cm}
\end{figure}

The first set of experiments evaluates the accuracy of constrained inference on unattributed histograms.  
We compare $\SC$ to the baseline approach $\SS$.  Since $\ss = \SS(I)$ is likely to be inconsistent---out-of-order, non-integral, and possibly negative---we consider a second baseline technique, denoted $\SR$, which enforces consistency by sorting $\ss$ and rounding each count to the nearest non-negative integer.

We evaluate the performance of these estimators on three queries from the three datasets.  On \trace: the query returns the number of internal hosts to which each external host is connected ($\approx 65$K external hosts); On \socnet, the query returns the degree sequence of the graph ($\approx 11$K nodes).  On \search, the query returns the search frequency over a 3-month period of the top $20$K keywords; position $i$ in the answer vector is the number of times the $i^{th}$ ranked keyword was searched.

\eat{
The runtime for computing $\SC$ varied across datasets and $\epsilon$, but the longest query took less than 3 minutes to compute (\trace, $\epsilon = 0.1$).
}

To evaluate the utility of an estimator, we measure its squared error.  Results report the average squared error over 50 random samples from the differentially-private mechanism (each sample produces a new $\ss$).  We also show results for three settings of $\epsilon = \{ 1.0, 0.1, 0.01 \}$; smaller $\epsilon$ means more privacy, hence more random noise.

Fig~\ref{fig:error_log} shows the results of the experiment.  Each bar represents average performance for a single combination of dataset, $\epsilon$, and estimator.  The bars represent, from left-to-right, $\SS$ (light gray), $\SR$ (gray), and $\SC$ (black).  The vertical axis is average squared error on a log-scale.  The results indicate that the proposed approach reduces the error by at least an order of magnitude across all datasets and settings of $\epsilon$.  Also, the difference between $\SR$ and $\SC$ suggests that the improvement is due not simply to enforcing integrality and non-negativity but from the way consistency is enforced through constrained inference (though $\SC$ and $\SR$ are comparable on \socnet{} at large $\epsilon$).  Finally, the relative accuracy of $\SC$ improves with decreasing $\epsilon$ (more noise).  Appendix~\ref{sec:app_unattributed} provides intuition for how $\SC$ reduces error.



\subsection{Universal Histograms} 
\label{expt:hierarchical_histograms}

\begin{figure}[!t]
\begin{center}
\includegraphics[width=3.5in]{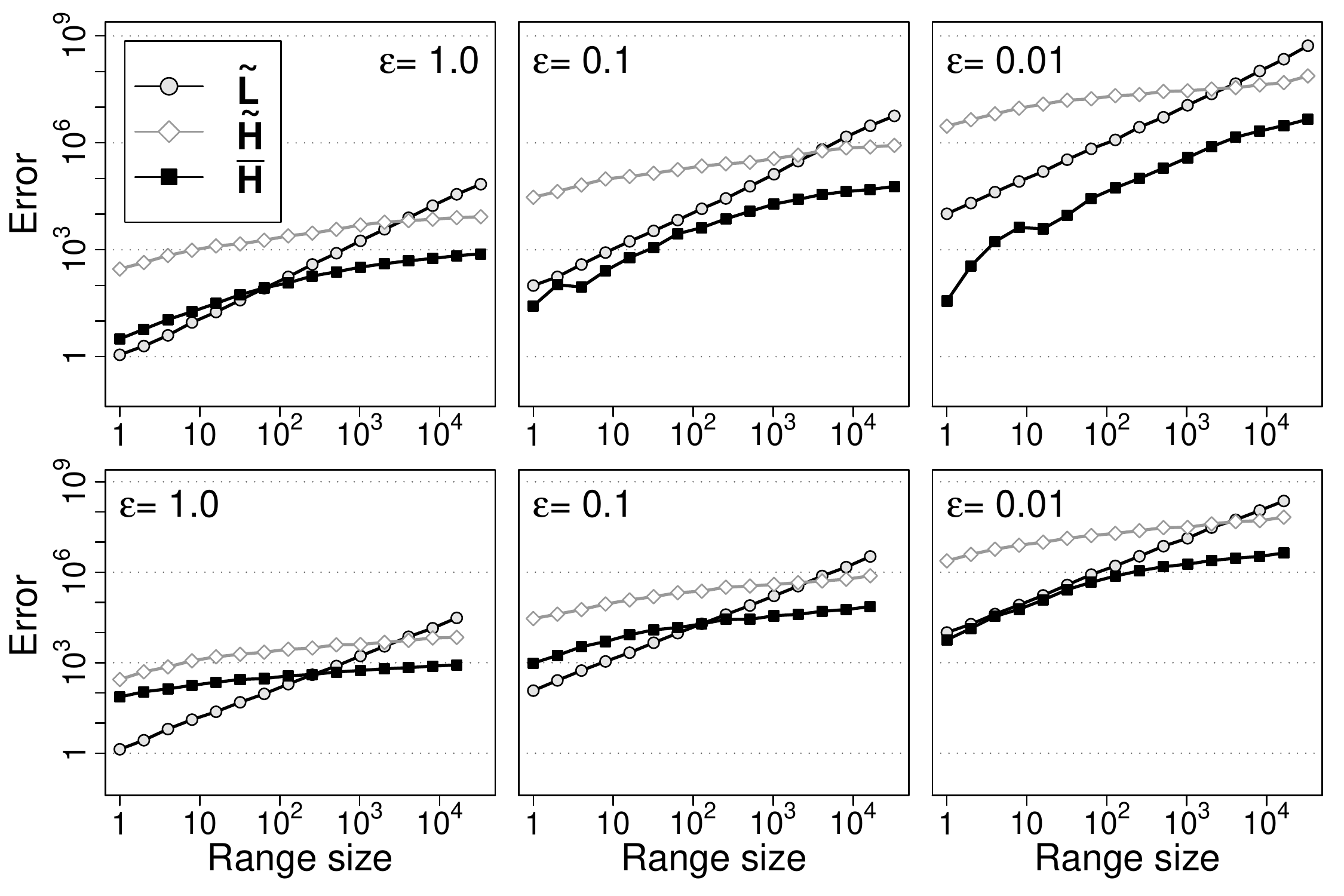}
\caption{
\label{fig:range_epsilon} A comparison estimators $\LL$ (circles), $\HH$ (diamonds), and $\HC$ (squares) on two real-world datasets (top \trace, bottom \search).
}
\end{center}
\vspace{-0.7cm}
\end{figure}

We now evaluate the effectiveness of constrained inference for the more general task of computing a universal histogram and arbitrary range queries.  
%
We evaluate three techniques for supporting universal histograms $\LL$, $\HH$, and $\HC$.
%
%
%
%
For all three approaches, we enforce integrality and non-negativity by rounding to the nearest non-negative integer.  With $\HC$, rounding is done as part of the inference process, using the approach described in Sec~\ref{sec:hiearchical_utility}.

We evaluate the accuracy over a set of range queries of varying size and location.  The range sizes are $2^i$ for $i=1, \dots, \hght-2$ where $\hght$ is the height of the tree.  For each fixed size, we select the location uniformly at random.  We report the average error over 50 random samples of $\tilde{l}$ and $\hh$, and for each sample, 1000 randomly chosen ranges.  

We evaluate the following histogram queries:  On \trace, the number of connections for each external host.  This is similar to the query in Sec~\ref{expt:unattributed_histograms} except that here the association between IP address and count is retained. On \search, the query reports the temporal frequency of the query term ``Obama'' from Jan. 1, 2004 to present.  (A day is evenly divided into 16 units of time.)
 
Fig~\ref{fig:range_epsilon} shows the results for both datasets and varying $\epsilon$.  The top row corresponds to \trace, the bottom to \search.  Within a row, each plot shows a different setting of $\epsilon \in \{ 1.0, 0.1, 0.01\}$.  For all plots, the x-axis is the size of the range query, and the y-axis is the error, averaged over sampled counts and intervals.  Both axes are in log-scale.




First, we compare $\LL$ and $\HH$.  For unit-length ranges, $\LL$ yields more accurate estimates.  This is unsurprising since it is a lower sensitivity query and thus less noise is added for privacy.  However, the error of $\LL$ increases linearly with the size of the range.  The average error of $\HH$ increases slowly with the size of the range, as larger ranges typically require summing a greater number of subtrees.  For ranges larger than about 2000 units, the error of $\LL$ is higher than $\HH$; for the largest ranges, the error of $\LL$ is 4-8 times larger than that of $\HH$ (note the log-scale).
	
Comparing $\HC$ against $\HH$, the error of $\HC$ is uniformly lower across all range sizes, settings of $\epsilon$, and datasets.  
The relative performance of the estimators depends on $\epsilon$.  At smaller $\epsilon$, the estimates of $\HC$ are more accurate relative to $\HH$ and $\LL$.  Recall that as $\epsilon$ decreases, noise increases.  This suggests that the relative benefit of statistical inference increases with the uncertainty in the observed data.
	
Finally, the results show that $\HC$ can have lower error than $\LL$ over small ranges, even for leaf counts.  This may be surprising since for unit-length counts, the scale of the noise of $\HC$ is \emph{larger} than that of $\LL$ by a factor of $\log n$.  The reduction in error is because these histograms are sparse.  When the histogram contains sparse regions, $\HC$ can effectively identify them because it has noisy observations at higher levels of the tree.  In contrast, $\LL$ has only the leaf counts; thus, even if a range contains no records, $\LL$ will assign a positive count to roughly half of the leaves in the range.

%

\section{Related work}
\label{sec:related}

Dwork has written comprehensive reviews of differential privacy~\cite{dwork2008differential,dwork2010a-firm}; below we highlight results closest to this work.

The idea of post-processing the output of a differentially private mechanism to ensure consistency was introduced in 
Barak et al.~\cite{barak2007privacy}, who proposed a linear program for making a set of marginals consistent, non-negative, and integral.  However, unlike the present work, the post-processing is not shown to improve accuracy.  



Blum et al.~\cite{blum2008a-learning} propose an efficient algorithm to publish synthetic data that is useful for range queries.  In comparison with our hierarchical histogram, the technique of Blum et al. scales slightly better (logarithmic versus poly-logarithmic) in terms of domain size (with all else fixed).  However, our hierarchical histogram achieves lower error for a fixed domain, and importantly, the error does not grow as the size of the database increases, whereas with Blum et al. it grows with $O(N^{2/3})$ with $N$ being the number of records (details in Appendix~\ref{app:blr}).  

The present work first appeared as an arXiv preprint~\cite{hay2009boosting0}, and since then a number of related works have emerged, including additional work by the authors.  The technique for unattributed histograms has been applied to accurately and efficiently estimate the degree sequences of large social networks~\cite{hay2009accurate}.  Several techniques for histograms over hierarchical domains have been developed.  
Xiao et al.~\cite{xiao2010differential} propose an approach based on the Haar wavelet, which is conceptually similar to the $\H$ query in that it is based on a tree of queries where each level in the tree is an increasingly fine-grained summary of the data.  In fact, that technique has error equivalent to a binary $\H$ query, as shown by Li et al.~\cite{li2010optimizing}, who represent both techniques as applications of the matrix mechanism, a framework for computing workloads of linear counting queries under differential privacy.
We are aware of ongoing work by McSherry et al.~\cite{mcsherry2009maximum} that combines  hierarchical querying with statistical inference, but differs from $\H$ in that it is adaptive. 
Chan et al.~\cite{chan2010private} consider the problem of continual release of aggregate statistics over streaming private data, and propose a differentially private counter that is similar to $\H$, in which items are hierarchically aggregated by arrival time.
%
The $\H$ and wavelet strategy are specifically designed to support range queries.  Strategies for answering more general workloads of queries under differential privacy are emerging, in both the offline~\cite{Hardt:2009On-the-Geometry-of-Differential,li2010optimizing} and online~\cite{roth2010interactive} settings.



\section{Conclusions}

Our results show that by transforming a differentially-private output so that it is consistent, we can boost accuracy.  Part of the innovation is devising a query set so that useful constraints hold.  Then the challenge is to apply the constraints by searching for the closest consistent solution.  Our query strategies for histograms have closed-form solutions for efficiently computing a consistent answer.  \eat{To our knowledge, they are the most accurate techniques for answering histogram queries under differential privacy.}  

Our results show that conventional differential privacy approaches can add more noise than is strictly required by the privacy condition.  We believe that using constraints may be an important part of finding optimal strategies for query answering under differential privacy.  More discussion of the implications of our results, and possible extensions, is included in Appendix~\ref{sec:discussion}.

\section{Acknowledgments}
Hay was supported by the Air Force Research Laboratory (AFRL) and IARPA, under agreement number FA8750-07-2-0158.  Hay and Miklau were supported by NSF CNS 0627642, NSF DUE-0830876, and NSF IIS 0643681.  Rastogi and Suciu were supported by NSF IIS-0627585.  The U.S. Government is authorized to reproduce and distribute reprints for Governmental purposes notwithstanding any copyright notation thereon.  The views and conclusion contained herein are those of the authors and should not be interpreted as necessarily representing the official policies or endorsements, either expressed or implied, of the AFRL and IARPA, or the U.S. Government.

\bibliographystyle{abbrv}
\bibliography{paper}

\begin{thebibliography}{10}

\bibitem{barak2007privacy}
B.~Barak, K.~Chaudhuri, C.~Dwork, S.~Kale, F.~McSherry, and K.~Talwar.
\newblock Privacy, accuracy, and consistency too: A holistic solution to
  contingency table release.
\newblock In {\em PODS}, 2007.

\bibitem{barlow1972statistical}
R.~E. Barlow, D.~J. Bartholomew, J.~M. Bremner, and H.~D. Brunk.
\newblock {\em Statistical Inference Under Order Restrictions}.
\newblock John Wiley and Sons Ltd, 1972.

\bibitem{barlow1972the-isotonic}
R.~E. Barlow and H.~D. Brunk.
\newblock The isotonic regression problem and its dual.
\newblock {\em JASA}, 67(337):140--147, 1972.

\bibitem{blum2008a-learning}
A.~Blum, K.~Ligett, and A.~Roth.
\newblock A learning theory approach to non-interactive database privacy.
\newblock In {\em STOC}, 2008.

\bibitem{chan2010private}
T.-H.~H. Chan, E.~Shi, and D.~Song.
\newblock Private and continual release of statistics.
\newblock In {\em ICALP}, 2010.

\bibitem{chunglu}
F.~R.~K. Chung and L.~Lu.
\newblock Survey: Concentration inequalities and martingale inequalities.
\newblock {\em Internet Mathematics}, 2006.

\bibitem{dwork2008differential}
C.~Dwork.
\newblock Differential privacy: A survey of results.
\newblock In {\em TAMC}, 2008.

\bibitem{dwork2010a-firm}
C.~Dwork.
\newblock A firm foundation for private data analysis.
\newblock {\em CACM, To appear}, 2010.

\bibitem{dwork2006calibrating}
C.~Dwork, F.~McSherry, K.~Nissim, and A.~Smith.
\newblock Calibrating noise to sensitivity in private data analysis.
\newblock In {\em TCC}, 2006.

\bibitem{ghosh2009universally}
A.~Ghosh, T.~Roughgarden, and M.~Sundararajan.
\newblock Universally utility-maximizing privacy mechanisms.
\newblock In {\em STOC}, 2009.

\bibitem{Hardt:2009On-the-Geometry-of-Differential}
M.~Hardt and K.~Talwar.
\newblock On the geometry of differential privacy.
\newblock In {\em STOC}, 2010.

\bibitem{hay2009accurate}
M.~Hay, C.~Li, G.~Miklau, and D.~Jensen.
\newblock Accurate estimation of the degree distribution of private networks.
\newblock In {\em ICDM}, 2009.

\bibitem{hay2009boosting0}
M.~Hay, V.~Rastogi, G.~Miklau, and D.~Suciu.
\newblock Boosting the accuracy of differentially-private queries through
  consistency.
\newblock {\em CoRR}, abs/0904.0942, April 2009.

\bibitem{hwang1994confidence}
J.~T.~G. Hwang and S.~D. Peddada.
\newblock Confidence interval estimation subject to order restrictions.
\newblock {\em Annals of Statistics}, 1994.

\bibitem{li2010optimizing}
C.~Li, M.~Hay, V.~Rastogi, G.~Miklau, and A.~McGregor.
\newblock Optimizing histogram queries under differential privacy.
\newblock In {\em PODS}, 2010.

\bibitem{mcsherry2009privacy}
F.~McSherry.
\newblock Privacy integrated queries: An extensible platform for
  privacy-preserving data analysis.
\newblock In {\em SIGMOD}, 2009.

\bibitem{mcsherry2009maximum}
F.~McSherry, K.~Talwar, and O.~Williams.
\newblock Maximum likelihood data synthesis.
\newblock Manuscript, 2009.

\bibitem{Newman:2003The-Structure-and-Function}
M.~E.~J. Newman.
\newblock The structure and function of complex networks.
\newblock {\em SIAM Review}, 45(2):167--256, 2003.

\bibitem{Owen:1982:GT}
G.~Owen.
\newblock {\em Game Theory}.
\newblock Academic Press Ltd, 1982.

\bibitem{roth2010interactive}
A.~Roth and T.~Roughgarden.
\newblock Interactive privacy via the median mechanism.
\newblock In {\em STOC}, 2010.

\bibitem{gaussmarkov}
S.~D. Silvey.
\newblock {\em Statistical Inference}.
\newblock Chapman-Hall, 1975.

\bibitem{xiao2010differential}
X.~Xiao, G.~Wang, and J.~Gehrke.
\newblock Differential privacy via wavelet transforms.
\newblock In {\em ICDE}, 2010.

\end{thebibliography}

\appendix

\section{Notational Conventions}
\label{sec:conventions}

The table below summarizes notational conventions used in the paper.

\begin{table}[h] 
\vspace{-0.1cm}
\begin{center}
{\small
\begin{tabular}{lp{1.6in}}
$\Q$ & Sequence of counting queries \\  
$\L$ & Unit-{\bf L}ength query sequence \\
$\H$ & {\bf H}ierarchical query sequence \\
$\S$ & {\bf S}orted query sequence \\
$\gamma_{\Q}$ & Constraint set for query $\Q$ \\  
$\QQ,\LL,\HH,\SS$ & Randomized query sequence \\  
$\HC,\SC$ & Randomized query sequence,\quad\quad returning minimum $L_2$ solution \\
$I$ & Private database instance \\ 
$\L(\db),\H(\db),\S(\db)$ & Output sequence (truth) \\
$\lsmall=\LL(\db), \hh=\HH(\db), \ss=\SS(\db)$ & Output sequence (noisy)\\
$\hbar=\HC(\db),\sbar=\SC(\db)$ & Output sequence (inferred) \\
\end{tabular}
}
\end{center}
\label{tbl:notation}
\vspace{-0.5cm}
\end{table}%

\section{Discussion of Main Results}
\label{sec:discussion}

Here we provide a supplementary discussion of the results, review the insights gained, and discuss future directions. 

\paragraph*{Unattributed histograms} The choice of the sorted query $\S$, instead of $\L$, is an unqualified benefit, because we gain from the inequality constraints on the output, while the sensitivity of $\S$ is no greater than that of $\L$.  Among other applications, this allows for extremely accurate estimation of degree sequences of a graph, improving error by an order of magnitude on the baseline technique.  The accuracy of the estimate depends on the input sequence.  It works best for sequences with duplicate counts, which matches well the degree sequences of social networks encountered in practice.  

Future work specifically oriented towards degree sequence estimation could include a constraint enforcing that the output sequence is {\em graphical}, i.e. the degree sequence of some graph. 
\eat{we have described an algorithm $\SC$ that provides differential privacy while providing uniformly greater accuracy than $\SS$ and $\LL$.  
	\2 How can we achieve such improvements?  Intuitively, a degree sequence should be estimated accurately under DP?  Why?
}

\paragraph*{Universal histograms} The choice of the hierarchical counting query $\H$, instead of $\L$, offers a trade off because the sensitivity of $\H$ is greater than that of $\L$.  It is interesting that for some data sets and privacy levels, the effect of the $\H$ constraints outweighs the increased noise that must be added.  In other cases, the algorithms based on $\H$ provide greater accuracy for all but the smallest ranges.  We note that in many practical settings, domains are large and sparse.  The sparsity implies that no differentially private technique can yield meaningful answers for unit-length queries because the noise necessary for privacy will drown out the signal.  So while $\LL$ sometimes has higher accuracy for small range queries, this may not have practical relevance since the relative error of the answers renders them useless.

In future work we hope to extend the technique for universal histograms to multi-dimensional range queries, and to investigate optimizations such as higher branching factors.

\vspace{3ex}

Across both histogram tasks, our results clearly show that it is possible to achieve greater accuracy without sacrificing privacy. The existence of our improved estimators $\SC$ and $\HC$ show that there is another differentially private noise distribution that is more accurate than independent Laplace noise.  This does not contradict existing results because the original differential privacy work showed only that calibrating Laplace noise to the sensitivity of a query was {\em sufficient} for privacy, not that it was necessary.  Only recently has the optimality of this construction been studied (and proven only for single queries) \cite{ghosh2009universally}.  Finding the optimal strategy for answering a set of queries under differential privacy is an important direction for future work, especially in light of emerging private query interfaces \cite{mcsherry2009privacy}.

A natural goal is to describe directly the improved noise distributions implied by $\SC$ and $\HC$, and build a privacy mechanism that samples from it.  This could, in theory, avoid the inference step altogether.  But it is seems quite difficult to discover, describe, and sample these improved noise distributions, which will be highly dependent on a particular query of interest.  Our approach suggests that constraints and constrained inference can be an effective path to discovering new, more accurate noise distributions that satisfy differential privacy.  As a practical matter, our approach does not necessarily burden the analyst with the constrained inference process because the server can implement the post-processing step.  In that case it would appear to the analyst as if the server was sampling from the improved distribution.

While our focus has been on histogram queries, the techniques are probably not limited to histograms and could have broader impact.  However, a general formulation may be challenging to develop.  There is a subtle relationship between constraints and sensitivity: reformulating a query so that it becomes highly constrained may similarly increase its sensitivity.  A challenge is finding queries such as $\S$ and $\H$ that have useful constraints but remain low sensitivity.  Another challenge is the computational efficiency of constrained inference, which is posed here as a constrained optimization problem with a quadratic objective function.  The complexity of solving this problem will depend on the nature of the constraints and is NP-Hard in general.   Our analysis shows that the constraint sets of $\S$ and $\H$ admit closed-form solutions that are efficient to compute. 

\section{Additional Experiments}
\label{sec:experiments_appendix}

\begin{figure*}[hbt]
\vspace{-0.5cm}
\begin{center}
\includegraphics[width=6.0in]{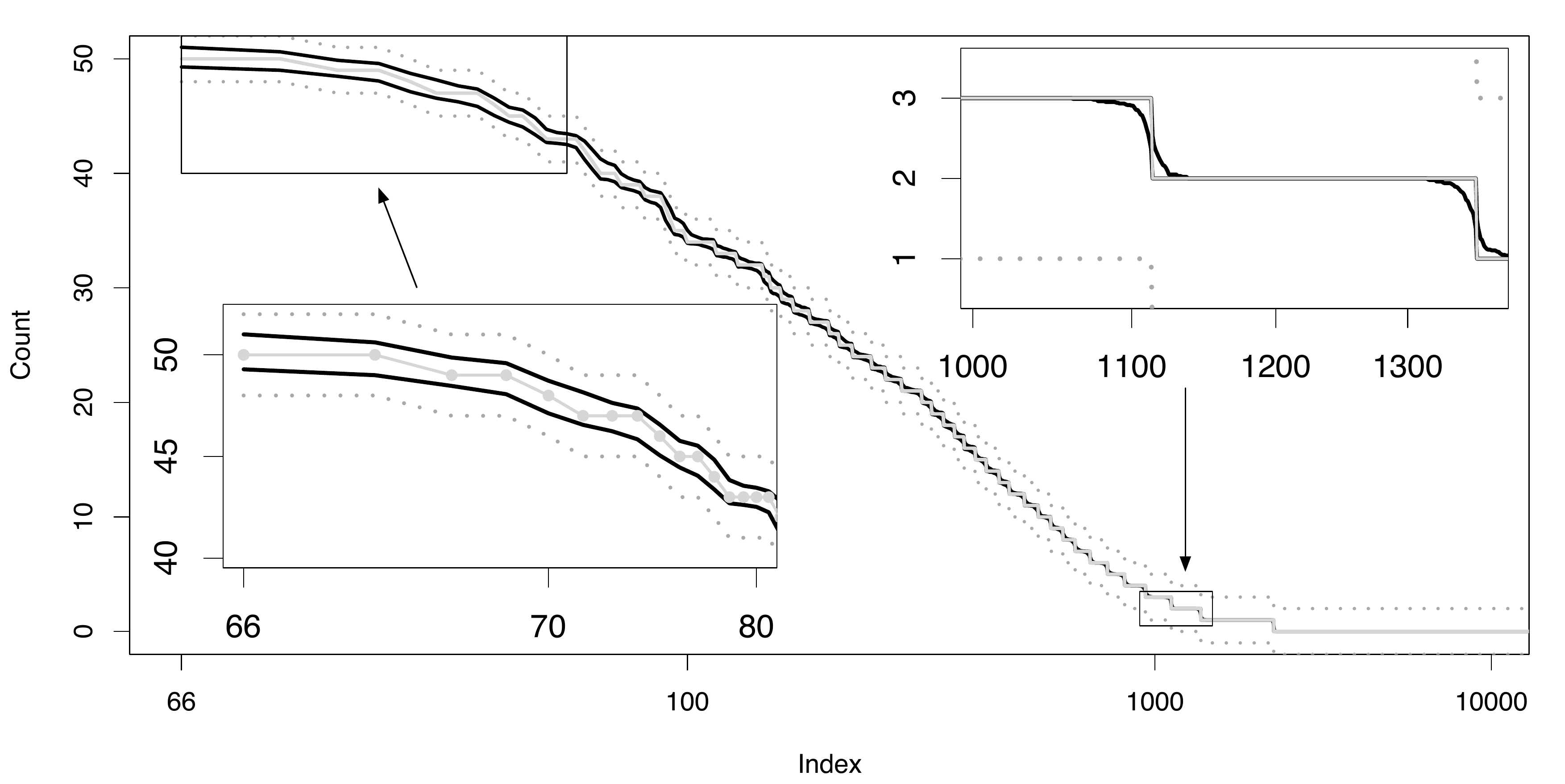}
\caption{
\label{fig:sequence_error} On \trace{}, $\S(I)$ (solid gray), the average error of $\SC$ (solid black) and $\SS$ (dotted gray), for $\epsilon=1.0$.
}
\end{center}
\vspace{-0.6cm}
\end{figure*}


This section provides detailed descriptions of the datasets, and additional results for unattributed histograms.

\trace{} is derived from an IP-level network trace collected at a major university.  The trace monitors traffic at the gateway between internal IP addresses and external IP addresses.  From this data, we derived a bipartite connection graph where the nodes are hosts, labeled by their IP address, and an edge connotes the transmission of at least one data packet.  Here, differential privacy ensures that individual connections remain private.

\socnet{} is a graph derived from friendship relations on an online social network site.  The graph is limited to a population of roughly $11,000$ students from a single university.  Differential privacy implies that friendships will not be disclosed.  The size of the graph (number of students) is assumed to be public knowledge.\footnote{This is not a critical assumption and, in fact, the number of students can be estimated privately within $\pm 1/\epsilon$ in expectation.  Our techniques can be applied directly to either the true count or a noisy estimate.}

\search{} is a dataset of search query logs over time from Jan. 1, 2004 to the present.  For privacy reasons, it is difficult to obtain such data.  Our dataset is derived from a search engine interface that publishes summary statistics for specified query terms.  We combined these summary statistics with a second dataset, which contains actual search query logs but for a much shorter time period, to produce a synthetic data set.  In the experiments, ground truth refers to this synthetic dataset.  \eat{(See \cite{barbaro2006a-face} for information on the privacy breaches that resulted from the publication of the ``anonymized'' query logs of~\cite{aol}.)} Differential privacy guarantees that the output will prevent the association of an individual entity (user, host) with a particular search term.

\paragraph*{Unattributed histograms}
\label{sec:app_unattributed}

Figure~\ref{fig:sequence_error} provides some intuition for how inference is able to reduce error.  Shown is a portion of the unattrib\-uted histogram of \trace: the sequence is sorted in \emph{descending} order along the x-axis and the y-axis indicates the count.  The solid gray line corresponds to ground truth: a long horizontal stretch indicates a subsequence of uniform counts and a vertical drop indicates a decrease in count.  The graphic shows only the middle portion of the unattributed histogram---some very large and very small counts are omitted to improve legibility.  The solid black lines indicate the error of $\SC$ averaged over 200 random samples of $\SS$ (with $\epsilon=1.0$); the dotted gray lines indicate the expected error of $\SS$.  

The inset graph on the left reveals larger error at the beginning of the sequence, when each count occurs once or only a few times.  However, as the counts become more concentrated (longer subsequences of uniform count), the error diminishes, as shown in the right inset.  Some error remains around the points in the sequence where the count changes, but the error is reduced to zero for positions in the middle of uniform subsequences.  

Figure~\ref{fig:sequence_error} illustrates that our approach reduces or eliminates noise in precisely the parts of the sequence where the noise is unnecessary for privacy.  Changing a tuple in the database cannot change a count in the middle of a uniform subsequence, only at the end points.  These experimental results also align with Theorem~\ref{theorem:unattributed}, which states that the error of $\SC$ is a function of the number of distinct counts in the sequence.  In fact, the experimental results suggest that the theorem also holds locally for subsequences with a small number of distinct counts.  This is an important result since the typical degree sequences that arise in real data, such as the power-law distribution, contain very large uniform subsequences.

\section{Proofs}





\newcommand{\inputSet}[1]{\mathcal{S}(#1)}
\def\qc{\qbar}
\begin{proof}[of Proposition~\ref{prop:inference_privacy}]
	For any output $\qc$, then let $\inputSet{\qc}$ denote the set of noisy answers such that if $\qq \in \inputSet{\qc}$ then the minimum $L_2$ solution given $\qq$ and $\cons_{\Q}$ is $\qc$.  For any $\db$ and $\db' \in \nbrs(\db)$, the following shows that $\QC$ is $\epsilon$-differentially private:
\begin{align*}
	Pr[ \QC(\db) = \qc ] &= Pr[ \QQ(\db) \in \inputSet{\qc} ] \\
	&\leq \exp(\epsilon) \; Pr[ \QQ(\db') \in \inputSet{\qc} ] \\
	&= \exp(\epsilon) \; Pr[ \QC(\db') = \qc ]
\end{align*}
where the inequality is because $\QQ$ is $\epsilon$-differentially private.
\end{proof}

\label{proof:sensitivity}


\begin{proof}[of Proposition~\ref{prop:sensitivity_s}]
		Given a database $\db$, suppose we add a record to it to obtain $\db'$.  The added record affects one count in $\L$, i.e., there is exactly one $i$ such that $\L(\db)[i] = x$ and $\L(\db')[i] = x + 1$, and all other counts are the same.  The added record affects $\S$ as follows.  Let $j$ be the largest index such that $\S(\db)[j] = x$, then the added record increases the count at $j$ by one: $\S(\db')[j] = x + 1$.  Notice that this change does not affect the sort order---i.e., 
	in $\S(\db')$, the $j^{th}$ value remains in sorted order: $\S(\db')[j-1] \leq x$, $\S(\db')[j] = x + 1$, and $\S(\db')[j+1] \geq x+1$.
All other counts in $\S$ are the same, and thus the $L_1$ distance between $\S(\db)$ and $\S(\db')$ is 1.
\end{proof}



\begin{proof}[of Proposition~\ref{prop:sensitivity_h}]
	If a tuple is added or removed from the relation, this affects the count for every range that includes it.  There are exactly $\hght$ ranges that include a given tuple: the range of a single leaf and the ranges of the nodes along the path from that leaf to the root.  Therefore, adding/removing a tuple changes exactly $\hght$ counts each by exactly 1.  Thus, the sensitivity is equal to $\hght$, the height of the tree.  
\end{proof}

%
%
%
%


\subsection{Proof of Theorem~\ref{thm:minL2OC}}
\label{app:proof_solution_unattributed}

We first restate the theorem below. Recall that $\ss[i,j]$ denotes the subsequence of $j-i+1$ elements: $\<\ss[i]$, $\ss[i+1]$, $\ldots, \ss[j]\>$.  Let $\mm[i,j]$ record the mean of these elements, i.e. $\mm[i,j] = \sum_{k=i}^j \ss[k] / (j-i+1)$. 

\minLTwoOC*

\begin{proof}
	In the proof, we abbreviate the notation and implicitly assume that the range of $i$ is $[1, n]$ or $[1, j]$ when $j$ is specified.  Similarly, the range of $j$ is $[1,n]$ or $[i,n]$ when $i$ is specified.
	
We start with the easy part, showing that $U_k \leq L_k$.  Define an $n \times n$ matrix
$A^k$ as follows:
\begin{eqnarray*}
A^k_{ij} & = &
\left\{
\begin{array}{ll}
\mm[i,j] & \mbox{ if i $\leq$ j} \\
\infty & \mbox{ if j $<$  i $\leq$ k} \\
-\infty & \mbox{otherwise}
\end{array}
\right.
\end{eqnarray*}
Then $\min_j \max_i A^k_{ij} = L_k$ and $\max_i \min_j A^k_{ij} = U_k$. In any
matrix $A^k$, $\max_i \min_j A^k_{ij} \leq \min_j \max_i A^k_{ij}$: this is a
simple fact that can be checked directly, or see~\cite{Owen:1982:GT},
hence $U_k \leq L_k$. 

We show next that if $\sbar$ is the minimum $L_2$ solution, then $L_k \leq
\sbar[k] \leq U_k$.  If we show this, then the proof of the theorem is completed, as then we will then have $\sbar[k] = L_k = U_k$.  The proof relies on the following lemma.

%

\begin{lemma} \label{lemma:bounds} Let $\sbar$ be the minimum $L_2$ solution. Then (i) $\sbar[1] \leq U_1$, (ii) $\sbar[n] \geq L_n$, (iii) for all $k$, $\min(\sbar[k+1], \max_{i} \mm[i,k]) \leq \sbar[k] \leq \max(\sbar[k-1], \min_{j} \mm[k,j])$.
\end{lemma}

The proof of the lemma appears below, but now we use it to complete the proof of Theorem~\ref{thm:minL2OC}.  First,  we show that $\sbar[k] \leq U_k$ using induction on $k$.  The base case is $k=1$ and it is stated in the lemma, part (i).  For the inductive step, assume $\sbar[k-1] \leq U_{k-1}$.  From (iii), we have that 
\begin{eqnarray*}
	\sbar[k] &\leq& \max ( \sbar[k-1], \min_j \mm[k,j] )\\ 
	& \leq & \max ( U_{k-1}, \min_j \mm[k,j] ) = U_k 
\end{eqnarray*}
The last step follows from the definition of $U_k$.  A similar induction argument shows that $\sbar[k] \geq L_k$, except the order is reversed: the base case is $k=n$ and the inductive step assumes $\sbar[k+1] \geq L_{k+1}$.
\end{proof}

The only remaining step is to prove the lemma.

\begin{proof}[of Lemma~\ref{lemma:bounds}] For (i), it is sufficient to prove that $\sbar[1] \leq \mm[1,j]$ for all $j \in [1,n]$. Assume the contrary. Thus there exists a $j$ such that for $\sbar[1] > \mm[1,j]$. Let $\delta = \sbar[1] - \mm[1,j]$. Thus $\delta > 0$. Further, for all $i$, denote $\delta_i = \sbar[i] - \sbar[1]$. Consider the sequence $\sbar'$ defined as follows:
\begin{eqnarray*}
\sbar'[i] & = &
\left\{
\begin{array}{ll}
\sbar[i] - \delta & \mbox{if $i \leq j$} \\
\sbar[i] & \mbox{otherwise}
\end{array}
\right.
\end{eqnarray*}
It is obvious to see that since $\sbar$ is a sorted sequence, so is $\sbar'$.

We now claim that $\Ltwo{\sbar'}{\ss} < \Ltwo{\sbar}{\ss}$. For this note that since the sequence $\sbar'[j+1,n]$ is identical to the sequence $\sbar[j+1,n]$, it is sufficient to prove $\Ltwo{\sbar'[1,j]}{\ss[1,j]} < \Ltwo{\sbar[1,j]}{\ss[1,j]}$.  To prove that, note that $\Ltwo{\sbar[1,j]}{\ss[1,j]}$ can be expanded as
\begin{eqnarray*}
 \Ltwo{\sbar[1,j]}{\ss[1,j]} &=& \sum_{i=1}^j (\sbar[i] - \ss[i])^2 = \sum_{i=1}^j (\sbar[1] + \delta_i - \ss[i])^2 \\
                             &=& \sum_{i=1}^j (\mm[1,j] + \delta + \delta_i - \ss[i])^2 
\end{eqnarray*}
Suppose for a moment that we fix $\mm[1,j]$ and $\delta_i$'s, and treat $\Ltwo{\sbar[1,j]}{\ss[1,j]}$ as a function $f$ over $\delta$.  The derivative of $f(\delta)$ is:
\begin{eqnarray*} 
  f'(\delta) &=& 2 \sum_{i=1}^j (\mm[1,j] + \delta + \delta_i - \ss[i]) \\
        &=& 2 \bigl(j \mm[1,j] - \sum_{i=1}^j \ss[i] \bigr)+ 2j\delta + 2 \sum_{i=1}^j \delta_i \\
&=& 2j\delta + 2\sum_{i=1}^j \delta_i
\end{eqnarray*} 
Since $\delta_i \geq 0$ for all $i$, then the derivative is strictly greater than zero for any $\delta > 0$, which implies that $f$ is a strictly increasing function of $\delta$ and has a minimum at $\delta = 0$.  Therefore, $\Ltwo{\sbar[1,j]}{\ss[1,j]} = f(\delta) > f(0) = \Ltwo{\sbar'[1,j]}{\ss[1,j]}$.  This is a contradiction since it was assumed that $\sbar$ was the minimum solution.  This completes the proof for (i).  \\
  
For (ii), the proof of $\sbar[n] \geq \max_i \mm[i,n]$ follows from a similar argument: if $\sbar[n] < \mm[i,n]$ for some $i$, define $\delta = \mm[i,n] - \sbar[n]$ and the sequence $\sbar'$ with elements $\sbar'[j] = \sbar[j] + \delta$ for $j \geq i$. Then $\sbar'$ can be shown to be a strictly better solution than $\sbar$, proving (ii). \\

For the proof of (iii), we first show that $\sbar[k] \leq \max(\sbar[k-1], \min_j \mm[k,j])$. Assume the contrary, i.e. there exists a $k$ such that $\sbar[k] > \sbar[k-1]$ and $\sbar[k] > \min_j \mm[k,j]$. In other words, we assume there exists a $k$ and $j$ such that $\sbar[k] > \sbar[k-1]$ and $\sbar[k] > \mm[k,j]$. Denote $\delta = \sbar[k] - \max(\sbar[k-1],\mm[k,j])$. By our assumption above, $\delta > 0$. Define the sequence

\begin{eqnarray*}
\sbar'[i] & = &
\left\{
\begin{array}{ll}
\sbar[i] - \delta & \mbox{ if $k \leq i \leq j$} \\
\sbar[i] & \mbox{otherwise}
\end{array}
\right.
\end{eqnarray*}

Note that by construction, $\sbar'[k] = \sbar[k] - \delta = \sbar[k] - (\sbar[k] - \max(\sbar[k-1],\mm[k,j])) = \max(\sbar[k-1],\mm[k,j])$. It is easy to see that $\sbar'$ is sorted~(indeed the only inversion in the sort order could have occurred if $\sbar'[k-1] > \sbar'[k]$, but doesn't as $\sbar'[k-1] = \sbar[k-1] \leq \max(\sbar[k-1],\mm[k,j]) = \sbar'[k]$). 

Now a similar argument as in the proof of (i) for the sequence $\ss[k,j]$, yields that the error $\Ltwo{\sbar'[k,j]}{\ss[k,j]}  < \Ltwo{\sbar[k,j]}{\ss[k,j]}$. Thus $\Ltwo{\sbar'}{\ss} < \Ltwo{\sbar'}{\ss}$ and $\sbar'$ is a strictly better solution than $\sbar$. This yields a contradiction as $\sbar$ is the minimum $L_2$ solution. Hence $\sbar[k] \leq \max(\sbar[k-1], \min_j \mm[k,j])$.

A similar argument in the the reverse direction shows that $\sbar[k] \geq \min(\sbar_{k+1}, \max_i \mm[i,k])$ completing the proof of (iii). 
\end{proof}

\subsection{Proof of Theorem~\ref{theorem:unattributed}}
\label{app:proof_unattributed}

We first restate the theorem below. Denote $n$ and $d$ as the number of values and the number of distinct values in $\S(I)$ respectively. Let $n_1,n_2,\ldots,n_d$ be the number of times each of the $d$ distinct values occur in $\S(I)$~(thus $\sum_i n_i = n$).

\thmUnattributed*

Before showing the proof, we prove the following lemma.

\begin{lemma} \label{lemma:translation} Let $s=\S(I)$ be the input sequence. Call a translation of $s$ the operation of subtracting from each element of $s$ a fixed amount $\delta$. Then $\error(\SC[i])$ is invariant under translation for all $i$.
\end{lemma}
\begin{proof}
Denote $Pr(\sbar|s)$~($Pr(\ss|s)$) the probability that $\sbar$~($\ss$) is output on the input sequence $s$. Denote $s'$, $\sbar'$, and $\ss'$ the sequence obtained by translating $s$, $\sbar$, and $\ss$ by $\delta$, respectively.

First observe that $Pr(\ss|s) = Pr(\ss'|s')$ as $\ss$ and $\ss'$ are obtained by adding the same Laplacian noise to $s$ and $s'$, respectively. Using Theorem~\ref{thm:minL2OC}~(since all $U_k$'s and $L_k$'s shift by $\delta$ on translating $\ss$ by $delta$), we get that if $\sbar$ is the minimum $L_2$ solution given $\ss$, then $\sbar'$ is the minimum $L_2$ solution given $\ss'$. Thus, $Pr(\sbar|s) = Pr(\sbar'|s')$ for all sequences $\sbar$. Further, since $\sbar[i]$ and $\sbar'[i]$ yield the same $L_2$ error with $s[i]$ and $s'[i]$ respectively, we get that the expected $\error(\SC[i])$ is same for both inputs $s$ and $s'$.
\end{proof}

\begin{lemma} \label{lemma:integral} Let $X$ be any positive random variable that is bounded~($\lim_{x \rightarrow \infty} xPr(X > x)$ exists). Then 
\begin{eqnarray*}
\E(X) \leq \int_0^\infty \! Pr( X > x) dx
\end{eqnarray*}
\end{lemma}
\begin{proof} The proof follows from the following chain of equalities.

\begin{eqnarray*}
&& \E(X) = \int_0^\infty \! x\frac{\partial}{\partial x} \left( Pr(X \leq x) \right) \\
  &=& -\int_0^\infty \! x\frac{\partial}{\partial x} \left( Pr(X > x) \right)  \\
  &=& -[xPr(X > x)]_0^{\infty} + \int_0^\infty \! (Pr( X \leq x)-1) dx~~\mbox{(by parts)}\\
  &=& -\lim_{x \rightarrow \infty} xPr(X > x) + \int_0^\infty \! Pr( X > x) dx \\
  &\leq& \int_0^\infty \! Pr( X > x) dx        				
\end{eqnarray*}

Here the last equality follows as $X$ is bounded and therefore the limit exists and is positive.  This completes the proof.
\eat{
Here the last equality follows as $X$ is bounded. Indeed,
\begin{eqnarray*}
\lim_{x \rightarrow \infty} xPr(X > x) = \lim_{x \rightarrow \infty} \frac{dPr(X > x)}{dx} = 0
\end{eqnarray*}
This completes the proof.
}
\end{proof}

We next state a theorem that was shown in~\cite{chunglu}

\begin{theorem}[Theorem 3.4~\cite{chunglu}] \label{thm:concentration} Suppose that $X_1$, $X_2$, \ldots, $X_n$ are independent random variables satisfying $X_i \leq \E(X_i) + M$, for $1 \leq i \leq n$. We consider the sum $X = \sum_{i=1}^{n} X_i$ with expectation $\E(X) = \sum_{i=1}^{n} E(X_i)$ and $Var(X) = \sum_{i=1}^n Var(X_i)$. Then, we have
\begin{eqnarray*}
Pr( X \geq \E(X) + \lambda ) \leq e^{\frac{-\lambda^2}{2(Var(X) + M\lambda/3)}}
\end{eqnarray*}
\end{theorem}

For a random variable $X$, denote $\one{X}$ the indicator function that $X \geq 0$~(thus $\one{X} = 1$ if $X \geq 0$ and $0$ otherwise). Using Theorem~\ref{thm:concentration}, we prove the following lemma.

\begin{lemma} \label{lemma:bounds:means} Suppose $i,j$ are indices such that for all $k \in [i,j]$, $s[k] \leq 0$. Then there exists a constant $c$ such that for all $\tau \geq 1$ the following holds.
\begin{eqnarray*}
Pr \left( \mm[i,j]^2\one{\mm[i,j]} \geq c(\frac{\log^2{\left( (j-i+1)\tau\right) }}{(j-i+1)\epsilon^2}) \right) \leq \frac{1}{(j-i+1)^2\tau^2}
\end{eqnarray*}
\end{lemma}

\begin{proof} We apply Theorem~\ref{thm:concentration} on $\ss[k]$ for $k \in [i,j]$. First note that $\E(\ss[k]) = s[k] \leq 0$. Further $Var(\ss[k]) = \frac{2}{\epsilon^2}$ as $\ss[k]$ is obtained by adding Laplace noise to $s[k]$ which has this variance. We also know that $\ss[k] \geq M + s[k]$ happens with probability at most $e^{-\epsilon M}/2$.

For simplicity, call $n$ to be $j-i+1$. Denoting $X = \sum_{k \in [i,j]} \ss[k]$, we see that $\E(X) \leq 0$ and $Var(X) = \frac{2n}{\epsilon^2}$. Further, set $M = 3\log{(n\tau)/\epsilon}$. Denote $B$ the event that for some $k$, $\ss[k] \geq M + s[k]$. Thus $Pr(B) \leq ne^{-\epsilon M}/2 \leq \frac{1}{2n^2\tau^3}$. If $B$ does not happen, we know that $\ss[k] \leq M + s[k]$ for all $k \in [i,j]$. Thus we can then apply Theorem~\ref{thm:concentration} to get:
\begin{eqnarray*}
Pr \left( X \geq \E(X) + \lambda \right) &\leq& e^{\frac{-\lambda^2}{2(2n/\epsilon^2 + \lambda\log{(n\tau)}/\epsilon)}} + Pr(B) \\
&=& e^{\frac{-\lambda^2}{2(2n/\epsilon^2 + \lambda\log{(n\tau)}/\epsilon)}} + \frac{1}{2n^2\tau^3}
\end{eqnarray*}
Setting $\lambda=\frac{8}{\epsilon}\sqrt{n}\log{(n\tau)}$ gives us that
\begin{eqnarray*}
Pr \left( X \geq \E(X) + \frac{8}{\epsilon}\sqrt{n}\log{(n\tau)} \right) &\leq& \frac{1}{n^2\tau^2}
\end{eqnarray*}
Since $\E(X) \leq 0$, we get
\begin{eqnarray*}
Pr \left( X \geq \frac{8}{\epsilon}\sqrt{n}\log{(n\tau)} \right) &\leq& \frac{1}{n^2\tau^2}
\end{eqnarray*}
Also we observe that $\mm[i,j] = X/n$, which yields
\begin{eqnarray*}
Pr \left( \mm[i,j] \geq \frac{8\log{(n\tau)}}{\sqrt{n}\epsilon} \right) &\leq& \frac{1}{n^2\tau^2}
\end{eqnarray*}
Finally, observe that $\mm[i,j] \leq c$ implies that $\mm[i,j]^2\one{\mm[i,j]} \leq c^2$. Thus we get
\begin{eqnarray*}
Pr \left( \mm[i,j]^2\one{mm[i,j]} \geq \frac{64\log^2{(n\tau)}}{n\epsilon^2} \right) &\leq& \frac{1}{n^2\tau^2}
\end{eqnarray*}
Putting $n = j-i+1$ and using $c=64$ gives us the required result.
\end{proof}

\eat{
\begin{lemma} \label{lemma:mean:variance} If $\m[i,j]=0$, then (i) $\E(\mm[i,j]^2) = O(\frac{1}{j-i+1})$, and (ii) $ Var(\mm[i,j]^2) = O(\frac{1}{(j-i+1)^2})$ 
\end{lemma}
\begin{proof}
Observe that a $Laplacian(2/\epsilon)$ random variable has a variance of $8/\epsilon^2$. Thus, the mean $\mm[i,j]$ of $j-i+1$ laplacian random variables has a variance of $\frac{8}{(j-i+1)\epsilon^2}$. By definition of variance, we get
\begin{eqnarray*} 
\E(\mm[i,j]^2)-E(\mm[i,j])^2 = \frac{8}{(j-i+1)\epsilon^2}\\
\E(\mm[i,j]^2)= \frac{8}{(j-i+1)\epsilon^2}
\end{eqnarray*} 
The second equation above follows from the first simply because $\E(\mm[i,j])=\m[i,j]=0$. Finally, (i) follows from the second equation. For (ii), by the definition of variance, we have
\begin{eqnarray*} 
Var(\mm[i,j]^2) &=& \E(\mm[i,j]^4)-E(\mm[i,j]^2)^2 \\
                &=& O(1/(j-i+1)^2)
\end{eqnarray*} 
The second equation above can be verified by expanding $\mm[i,j]^4$ in terms of the elements of $\ss$ and bounding the expectation of the terms in expansion. This completes the proof.
\end{proof}

\begin{lemma} \label{lemma:bounds:means} If $\m[i,j] \leq 0$ then there exists a constant $c$ depending only on $\epsilon$ such that for all $\tau \geq 0$ the following holds.
\begin{eqnarray*}
Pr \left( \mm[i,j]^2\one{\mm[i,j]} \geq c(\frac{1}{j-i+1} + \tau \frac{1}{(j-i+1)^2}) \right) \leq 1/\tau^2
\end{eqnarray*}
\end{lemma}

\begin{proof}

We first prove the statement for $\m[i,j]=0$. By Chebychev's inequality, we know that 
\begin{eqnarray}
Pr \left( \mm[i,j]^2 - \E(\mm[i,j]^2) \geq \tau Var(\mm[i,j]^2) \right) \leq 1/\tau^2 \\
\Rightarrow Pr \left( \mm[i,j]^2 \geq \E(\mm[i,j]^2) + \tau Var(\mm[i,j]^2) \right) \leq 1/\tau^2 \label{chebychev}
\end{eqnarray}

Since, from Lemma~\ref{lemma:mean:variance}, we have $\E(\mm[i,j]^2) = O(\frac{1}{j-i+1})$ and $Var(\mm[i,j]^2) = O(\frac{1}{(j-i+1)^2})$. Thus there exists a constant $c$ such that 
\begin{eqnarray}
E(\mm[i,j]^2) + \tau Var(\mm[i,j]^2) \leq c(\frac{1}{j-i+1} + \tau \frac{1}{(j-i+1)^2}) \label{Onotation}
\end{eqnarray}
Combining Eq~\ref{chebychev} and Eq~\ref{Onotation}, we get 
\begin{eqnarray*}
Pr \left( \mm[i,j]^2 \geq c(\frac{1}{j-i+1} + \tau \frac{1}{(j-i+1)^2}) \right) \leq 1/\tau^2
\end{eqnarray*}
Observing that $\mm[i,j]^2 \geq \mm[i,j]^2\one{\mm[i,j]}$ yields the required result. This completes the proof for the case $\m[i,j]=0$.

Next we consider the case when $\m[i,j] < 0$. Then define a sequence $s'$ by translating $s$ by $\m[i,j]$. Then $\mm'[i,j]$ the mean of elements $\ss'[i],..,\ss'[j]$ is $\mm[i,j] - \m[i,j]$. Since $\m[i,j] < 0$, we get $\mm'[i,j] > \mm[i,j]$. Further $\E(\mm'[i,j]) = \m[i,j] - \m[i,j] = 0$. Thus, we can apply the above proven result on $\mm'[i,j]$, which yields:
\begin{eqnarray*}
Pr \left( \mm'[i,j]^2\one{\mm'[i,j]} \geq c(\frac{1}{j-i+1} + \tau \frac{1}{(j-i+1)^2}) \right) \leq 1/\tau^2
\end{eqnarray*}

Finally, since $\mm'[i,j] > \mm[i,j]$ we conclude that $\mm'[i,j]^2\one{\mm[i,j]} \geq \mm[i,j]^2\one{\mm[i,j]}$. This yields the required result for $\m[i,j] < 0$ and completes our proof.
\end{proof}
}

Now we can give the proof of Theorem~\ref{theorem:unattributed}.

\begin{proof}[of Theorem~\ref{theorem:unattributed}] 
The proof of $\error(\SS) = \Theta(n/\epsilon^2)$ is obvious since:
\begin{eqnarray*}
\error(\SS) = \sum_{k=1}^n \error(\ss[i]) = n (\frac{2}{\epsilon^2})
\end{eqnarray*}

In the rest of the proof, we shall show bound $\error(\SC)$. Let $s=S(I)$ be the input sequence. We know that $s$ consists of $d$ distinct elements. Denote $s_r$ as the $r^{th}$ distinct element of $s$. Also denote $[l_r,u_r]$ as the set of indices corresponding to $s_r$, i.e. $\forall_{i \in [l_r,u_r]} s[i] = s_r$ and $\forall_{i \notin [l_r,u_r]} s[i] \neq s_r$. Let $\m[i,j]$ record the mean of elements in $s[i,j]$, i.e. $\m[i,j] = \sum_{k=i}^j  s[k] / (i-j+1)$. 
 
To bound $\error(\SC)$, we shall bound $\error(\SC[i])$ separately for each $i$. To bound $\error(\SC[i])$, we can assume W.L.O.G that $s[i]$ is $0$. This is because if $s[i] \neq 0$, then we can translate the sequence $s$ by $s[i]$. As shown in Lemma~\ref{lemma:translation} this preserves $\error(\SC[i])$, while making $s[i]=0$.

Let $k \in [l_r,u_r]$ be any index for the $r^{th}$ distinct element of $s$. By definition, $\error(\SC[k]) = \E(\sbar[k]-s[k])^2 = \E(\sbar[k]^2)$~(as we can assume W.L.O.G $s[k]=0$). From Theorem~\ref{thm:minL2OC}, we know that $\sbar[k] = U_k$. Thus  $\error(\SC[k]) = \E (U_k^2)$. Here we treat $U_k = max_{i \leq k} min_{j} \mm[i,j]$ as a random variable. Now by definition of $\E$, we have
\begin{eqnarray*}
\E (U_k^2) = \E (U_k^2 \one{U_k}) + \E (U_k^2 (1-\one{U_k}))
           = A+B~\mbox{(say)}           											
\end{eqnarray*}

\eat{
\begin{eqnarray*}
&& \E (U_k^2) = \int_0^\infty \! x\frac{\partial \left( Pr(U_k^2 \leq x) \right) }{\partial x} \\
   &=& \int_0^\infty \! x\frac{\partial \left( Pr(U_k^2 \leq x \wedge U_k \geq 0) + Pr(U_k^2 \leq x \wedge U_k \leq 0) \right) }{\partial x} \\
           &=& A+B~\mbox{(say)}           											
\end{eqnarray*}
}

We shall bound $A$ and $B$ separately. For bounding $A$, denote $\U_k = max_{i \leq k} \mm[i,u_r]$. It is apparent that $\U_k \geq U_k$ and thus $\U_k^2 \one{\U_k} \geq U_k^2 \one{U_k}$. To bound $A$, we observe that 

\begin{eqnarray*}
  A &=& \E (U_k^2 \one{U_k})
  \leq \E (\U_k^2 \one{\U_k}) 
\end{eqnarray*}

Further, since $\U_k = max_{i \leq k} \mm[i,u_r]$, we know that $\U_k^2\one{\U_k} = max_{i \leq k} \mm[i,u_r]^2\one{\mm[i,u_r]}$. Thus we can write:

\begin{eqnarray*}
 A &\leq& \E (\U_k^2 \one{\U_k}) = \E \left( max_{i \leq k} \mm[i,u_r]^2\one{\mm[i,u_r]} \right)
\end{eqnarray*}


\eat{
Here the final equality follows from Lemma~\ref{lemma:integral} as $\U_k\one{\U_k}$ is bounded. To see why it is bounded, it is enough to observe that each $\ss[i]$ is bounded. It follows that all linear combinations (i.e. $\mm[i,j]$) of elements of $\ss$ are bounded, and all quantities obtained by taking min/max~(i.e. $\U_k$) of $\mm[i,j]$'s are bounded.
}

\eat{
Let us denote $e[x]$ the event: $\U_k^2 > x  \wedge \U_k \geq 0$. Recall that $\U_k = max_{i \leq k} \mm[i,r_k]$. Denote $e_i[x]$ the event: $\mm[i,u_r]^2 > x  \wedge \mm[i,u_r] \geq 0$. Then it is easy to see that $e[x] \Rightarrow \vee_{i \leq k} e_i[x]$: since if $e[x]$ is true, we know that for $j = argmax_{i\leq k} M[i,u_r]$, $\mm[j,u_r]^2 > x$ and $\mm[j,u_r] \geq 0$. Thus, we get $Pr\left(e[x] \right) \leq Pr \left( \vee_{i\leq k} e_i[x] \right)$. Further, by union bound we know $Pr \left( \vee_{i\leq k} e_i[x] \right) \leq \sum_{i\leq k} Pr \left( e_i[x] \right)$, which yields:

\begin{eqnarray*}
A &\leq& \int_0^\infty \! Pr\left(\U_k^2 > x  \wedge \U_k \geq 0 \right) dx\\
  &=& \int_0^\infty \! Pr( e[x] )dx 
  \leq \int_0^\infty \! \sum_{i\leq k} Pr( e_i[x]) dx\\
  &=& \int_0^\infty \! \sum_{i\leq k} Pr \left( \mm[i,u_r]^2 > x  \wedge \mm[i,u_r] \geq 0 \right) dx
\end{eqnarray*}
}

Let $\tau > 1$ be any number and $c$ be the constant used in Lemma~\ref{lemma:bounds:means}. Let us denote $e_i$ the event that:
\begin{eqnarray*}
\mm[i,u_r]^2\one{\mm[i,u_r]} \geq c(\frac{\log^2{\left( (u_r-i+1)\tau\right) }}{(u_r-i+1)\epsilon^2})
\end{eqnarray*}

We can apply lemma~\ref{lemma:bounds:means} to compute the probability of $e_i$ as $\s[j] \leq 0$ for all $j \leq u_r$~(as we assumed W.L.O.G $s[k]=0$). Thus we get $Pr(e_i) \leq \frac{1}{(u_r-i+1)^2 \tau^2}$.

Define $e = \vee_{i=1}^{u_r} e_i$. Then $Pr(e) \leq \sum_{i=1}^{u_r} Pr(e_i) = 2/\tau^2$~(as $\sum_{i=1}^{u_r} 1/i^2 \leq 2$). If the event $e$ does not happen, then it is easy to see that

\begin{eqnarray*}
\U_k^2\one{\U_k} &=& max_{i \leq k} \mm[i,u_r]^2\one{\mm[i,u_r]} \\
       &\leq& c(\frac{\log^2{\left( (u_r-k+1)\tau\right) }}{(u_r-k+1)\epsilon^2})\\
\end{eqnarray*}

Thus with at least probability  $1-2/\tau^2$~(which is $Pr(\neg e)$), we get $\U_k^2\one{\U_k}$ is bounded as above. This yields that there exist constants $c_1$ and $c_2$ such that $\E(\U_k^2\one{\U_k}) \leq \frac{c_1\log^2{(u_r-k+1)} + c_2}{(u_r-k+1)\epsilon^2}$. The proof is by the application of Lemma~\ref{lemma:integral}~(as $\U_k$ is bounded) and a simple integration over $\tau$ ranging from $1$ to $\infty$. Finally we get that $A \leq \E(\U_k^2\one{\U_k}) \leq \frac{c_1\log^2{(u_r-k+1)} + c_2}{(u_r-k+1)\epsilon^2}$.

Recall that $B=\E (U_k^2 (1-\one{U_k}))$. We can write $B$ as $\E (L_k^2 (1-\one{L_k}))$ as $L_k = U_k$. Using the exact same arguments as above for $L_k$ but on sequence $-\S$ yields that $B \leq \frac{c_1\log^2{(k-l_r+1)} + c_2}{(k-l_r+1)\epsilon^2}$.

Finally, we get that $\SC[k] = A+B$ which is less than $\frac{c_1\log^2{(u_r-k+1)} + c_2}{(u_r-k+1)\epsilon^2}$ $+$ $\frac{c_1\log^2{(k-l_r+1)} + c_2}{(k-l_r+1)\epsilon^2}$.

To obtain a bound on the total $error(\SC)$.
\begin{eqnarray*}
error(\SC) &=& \sum_{r=1}^{d} \sum_{k \in [l_r,u_r]} error(\SC[k]) \\
&\leq& \sum_{r=1}^{d} \sum_{k \in [l_r,u_r]} \frac{c_1\log^2{(u_r-k+1)} + c_2}{(u_r-k+1)\epsilon^2} +\\
&&\sum_{r=1}^{d} \sum_{k \in [l_r,u_r]}  \frac{c_1\log^2{(k-l_r+1)} + c_2}{(k-l_r+1)\epsilon^2} \\
&\leq& \sum_{r=1}^d  \frac{c_1\log^3{(u_r-l_r+1)} + c_2}{\epsilon^2} 
\end{eqnarray*}

Finally noting that $u_r-l_r+1$ is just $n_r$, the number of occurrences of $s_r$ in $s$, we get $error(\SC) = \sum_r \frac{c_1\log^3{n_r}+c_2}{\epsilon^2} = O(d\log^3{n}/\epsilon^2)$. This completes the proof of the theorem.
\end{proof}


\subsection{Proof of Theorem~\ref{minL2HC}}
\label{app:proof_prop2}

%
%

We first restate the theorem below. 

\minLTwoHC*

\begin{proof} We first show that $\hbar[r] = z[r]$ for the root node $r$. By definition of a minimum $L_2$ solution, the sequence $\hbar$ satisfies the following constrained optimization problem.   Let $\Z[u] = \sum_{w \in succ(u)} z[w]$.

\begin{eqnarray*}
\mbox{minimize~} \sum_{v} (\hbar[v] - \hh[v])^2 \\
\mbox{subject to~} \forall v, \sum_{u \in succ(v)} \hbar[u] = \hbar[v]
\end{eqnarray*}

Denote $leaves(v)$ to be the set of leaf nodes in the subtree rooted at $v$. The above optimization problem can be rewritten as the following unconstrained minimization problem.

\begin{eqnarray*}
\mbox{minimize~} \sum_{v} \bigl( (\sum_{l \in leaves(v)} \hbar[l]) - \hh[v] \bigr)^2
\end{eqnarray*}

For finding the minimum, we take derivative w.r.t $\hbar[l]$ for each $l$ and equate it to $0$. We thus get the following set of equations for the minimum solution.

\begin{eqnarray*}
\forall l, \sum_{v: l \in leaves(v)} 2\bigl( (\sum_{l' \in leaves(v)} \hbar[l']) - \hh[v] \bigr) = 0
\end{eqnarray*}


Since $\sum_{l' \in leaves(v)} \hbar[l'] = \hbar[v]$, the above set of equations can be rewritten as:
$
\forall l, \sum_{v: l \in leaves(v)} \hbar[v] = \sum_{v: l \in leaves(v)} \hh[v] \label{equation:leaf}
$

For a leaf node $l$, we can think of the above equation for $l$ as corresponding to a path from $l$ to the root $r$ of the tree. The equation states that sum of the sequences $\hbar$ and $\hh$ over the nodes along the path are the same. We can sum all the equations to obtain the following equation.

\begin{eqnarray*}
\sum_{v} \sum_{l \in leaves(v)} \hbar[v] = \sum_{v} \sum_{l \in leaves(v)} \hh[v]
\end{eqnarray*}

Denote $level(i)$ as the set of nodes at height $i$ of the tree. Thus root belongs to $level(\hght-1)$ and leaves in $level(0)$. Abbreviating $LHS$~($RHS$) for the left~(right) hand side of the above equation, we observe the following.

\begin{eqnarray*}
LHS &=& \sum_{v} \sum_{l \in leaves(v)} \hbar[v]  \\ 
   &=& \sum_{i=0}^{\hght-1} \sum_{v \in level(i)} \sum_{l \in leaves(v)} \hbar[v] \\
   &=& \sum_{i=0}^{\hght-1} \sum_{v \in level(i)} k^i \hbar[v] = \sum_{i=0}^{\hght-1} k^i \sum_{v \in level(i)} \hbar[v] \\
   &=& \sum_{i=0}^{\hght-1} k^i \hbar[r] = \frac{k^\hght-1}{k-1} \hbar[r]
\end{eqnarray*}

Here we use the fact that $\sum_{v \in level(i)} \hbar[v] = \hbar[r]$ for any level $i$. This is because $\hbar$ satisfies the constraints of the tree. In a similar way, we also simplify the RHS.

\begin{eqnarray*}
RHS &=& \sum_{v} \sum_{l \in leaves(v)} \hh[v]  \\ 
   &=& \sum_{i=0}^{\hght-1} \sum_{v \in level(i)} \sum_{l \in leaves(v)} \hh[v] \\
   &=& \sum_{i=0}^{\hght-1} \sum_{v \in level(i)} k^i \hh[v] = \sum_{i=0}^{\hght-1} k^i \sum_{v \in level(i)} \hh[v]
\end{eqnarray*}

Note that we cannot simplify the RHS further as $\hh[v]$ may not satisfy the constraints of the tree. Finally equating $LHS$ and $RHS$ we get the following equation.

\begin{eqnarray*}
\hbar[r] = \frac{k-1}{k^\hght-1}\sum_{i=0}^{\hght-1} k^i \sum_{v \in level(i)} \hh[v]
\end{eqnarray*}

Further, it is easy to expand $z[r]$ and check that

\begin{eqnarray*}
z[r] = \frac{k-1}{k^\hght-1}\sum_{i=0}^{\hght-1} k^i \sum_{v \in level(i)} \hh[v]
\end{eqnarray*}
Thus we get $\hbar[r] = z[r]$.  For nodes $v$ other than the $r$, assume that we have computed $\hbar[u]$ for $u=pred(v)$. Denote $H = \hbar[u]$. Once $H$ is fixed, we can argue that the value of $\hbar[v]$ will be independent of the values of $\hh[w]$ for any $w$ not in the subtree of $u$. 
\eat{
This is because, denoting $subtree(u)$ the nodes in the subtree of $u$ the $L_2$ minimization problem can be rewritten as:

\begin{eqnarray*}
&&\mbox{minimize~} \sum_{v \in subtree(u)} (\hbar[v] - \hh[v])^2 +  \sum_{w \in subtree(u)} (\hbar[w] - \hh[w])^2 \\
&&\mbox{subject to~} \\
&& \forall v \in subtree(u), \sum_{v' \in succ(v)} \hbar[v'] = \hbar[v] \\
&& \sum_{v \in succ(u)} \hbar[v] = H \\
&& \forall w \notin subtree(u), \sum_{w' \in succ(w)} \hbar[w'] = \hbar[w]
\end{eqnarray*}
}

For nodes $w \in subtree(u)$ the $L_2$ minimization problem is equivalent to the following one.

\begin{eqnarray*}
&& \mbox{minimize~} \sum_{w \in subtree(u)} (\hbar[w] - \hh[w])^2\\
&& \mbox{subject to~} \forall w \in subtree(u), \sum_{w' \in succ(w)} \hbar[w'] = \hbar[w] \\
&& \mbox{and~} \sum_{v \in succ(u)} \hbar[v] = H
\end{eqnarray*}

Again using nodes $l \in leaves(u)$, we convert this minimization into the following one.

\begin{eqnarray*}
&&\mbox{minimize~} \sum_{w \in subtree(U)} \bigl( (\sum_{l \in leaves(w)} \hbar[l]) - \hh[w] \bigr)^2 \\
&&\mbox{subject to~} \sum_{l \in leaves(u)} \hbar[u] = H
\end{eqnarray*}

We can now use the method of Lagrange multipliers to find the solution of the above constrained minimization problem. Using $\lambda$ as the Lagrange parameter for the constraint $\sum_{l \in leaves(u)} \hbar[u] = H$, we get the following sets of equations.
 
\begin{eqnarray*}
\forall l \in leaves(u), \sum_{w: l \in leaves(w)} 2\bigl( \hbar[w] - \hh[w] \bigr) = -\lambda
\end{eqnarray*}

Adding the equations for all $l \in leaves(u)$ and solving for $\lambda$ we get $\lambda = -\frac{H - \Z[u]}{n(u)-1}$. Here $n(u)$ is the number of nodes in $subtree(u)$. Finally adding the above equations for only leaf nodes $l \in leaves(v)$, we get

\begin{eqnarray*}
\hbar[v] & = & z[v] -  (n(v)-1) \cdot \lambda \\ 
         & = & z[v] + \frac{n(v)-1}{n(u)-1} (H - \Z[u]])  \\
         & = & z[v] + \frac{1}{k}(\hbar[u] - \Z[u])
\end{eqnarray*}

This completes the proof.
\end{proof}


\subsection{Proof of Theorem~\ref{theorem:hierarchical:optimal}} 
\label{app:proof_thm4}

First, the theorem is restated.

\utilityH*


\begin{proof} 
	For (i), the linearity of $\HC$ is obvious from the definition of $z$ and $\hbar$. To show $\HC$ is unbiased, we first show that $z$ is unbiased, i.e. $\E (z[v]) = h[v]$. We use induction: the base case is if $v$ is a leaf node in which case $\E (z[v]) = \E(\hh[v]) = h[v]$. If $v$ is not a leaf node, assume that we have shown $z$ is unbiased for all nodes $u \in succ(v)$. Thus 
	\begin{eqnarray*}
	\E(\Z[v]) = \sum_{ u \in succ(v)} \E(z[u]) = \sum_{ u \in succ(v)} h[u] = h[v]
	\end{eqnarray*}
	Thus $\Z[v]$ is an unbiased estimator for $h[v]$. Since $z[v]$ is a linear combination of $\hh[v]$ and $\Z[v]$, which are both unbiased estimators, $z[v]$ is also unbiased. This completes the induction step proving that $z$ is unbiased for all nodes. Finally, we note that $\hbar[v]$ is a linear combination of $\hh[v]$, $z[v]$, and $\Z[v]$, all of which are unbiased estimators. Thus $\hbar[v]$ is also unbiased proving (i). 

	For (ii), we shall use the Gauss-Markov theorem~\cite{gaussmarkov}. We shall treat the sequence $\hh$ as the set of observed variables, and $l$, the sequence of original leaf counts, as the set of unobserved variables. It is easy to see that for all nodes $v$
	\begin{eqnarray*}
	 \hh[v] = \sum_{u \in leaves(v)} l[u] + noise(v)
	\end{eqnarray*}
	Here $noise(v)$ is the Laplacian random variable, which is independent for different nodes $v$, but has the same variance for all nodes. Hence $\hh$ satisfies the hypothesis of Gauss-Markov theorem. (i) shows that $\hbar$ is a linear unbiased estimator. Further, $\hbar$ has been obtained by minimizing the $L_2$ distance with $\hh[v]$. Hence, $\hbar$ is  the Ordinary Least Squares~(OLS) estimator, which by the Gauss-Markov theorem has the least $\error$.  Since it is the OLS estimator, it minimizes the $\error$ for estimating any linear combination of the original counts, which includes in particular the given range query $q$.
	

For (iii), we note that any query $q$ can be answered by summing at most $k \hght$ nodes in the tree. Since for any node $v$, $\error(\HC[v]) \leq \error(\HH[v]) = 2 \hght^2/\epsilon^2$, we get 
\begin{eqnarray*}
\error(\HC[q]) \leq  k \hght (2 \hght^2/\epsilon^2) = O(\hght^3/\epsilon^2)
\end{eqnarray*}

For (iv), denote $l_1$ and $l_2$ to be the leftmost and rightmost leaf nodes in the tree. Denote $r$ to be the root. We consider the query $q$ that asks for the sum of all leaf nodes except for $l_1$ and $l_2$. Then from (i) $\error(\HC(q))$ is less than the $\error$ of the estimate $\hh[r] - \hh[l_1] - \hh[l_2]$, which is $6 \hght^2/\epsilon^2$. But, on the other hand, $\HH$ will require summing $2(k-1)(\hght-1)-k$ noisy counts in total---$2(k-1)$ at each level of the tree, except for the root and the level just below the root, only $k-2$ counts are summed. Thus $\error(\HH_q) = 2(2(k-1)(\hght-1)-k)\hght^2/\epsilon^2$. Thus 
\begin{eqnarray*}
\error(\HC_q) \leq \frac{3\error(\HH_q)}{2(\hght-1)(k-1)-k}
\end{eqnarray*}
This completes the proof.
\end{proof}

\section{Comparison with Blum et al.}
\label{app:blr}

We compare a binary $\HH_q$ against the binary search equi-depth histogram of Blum et al.~\cite{blum2008a-learning} in terms of $(\epsilon, \delta)$-usefulness as defined by Blum et al.  Since $\epsilon$ is used in the usefulness definition, we use $\alpha$ as the parameter for $\alpha$-differential privacy.

Let $N$ be the number of records in the database.  An algorithm is $(\epsilon, \delta)$-useful for a class of queries if with probability at least $1 - \delta$, for every query in the class, the absolute error is at most $\epsilon N$.

For any range query $q$, the absolute error of $\HH_q$ is 
%
$| \HH_q(\db) - \H_q(\db) | = | Y |$
where $Y = \sum_{i=1}^c \gamma_i$, each $\gamma_i \sim Lap(\hght/\alpha)$, and $c$ is the number of subtrees in $\HH_q$, which is at most $2\hght$.  We use  Corollary 1 from ~\cite{chan2010private} to bound the error of a sum of Laplace random variables.  With $\nu = \sqrt{c \hght^2/\alpha^2} \sqrt{2 \ln \frac{2}{\delta'}}$, we obtain
\[
Pr\left[ | Y | \; \leq \; \frac{16 \hght^{\frac{3}{2}} \ln \frac{2}{\delta'}}{\alpha}  \right] \geq 1 - \delta'
\]

The above is for a single range query.  To bound the error for all $n \choose 2$ range queries, we use a union bound.  Set $\delta' = \frac{\delta}{n^2}$.  Then $\HH$ is $(\epsilon, \delta)$-useful provided that $\epsilon \geq \left( {16 \hght^{\frac{3}{2}}  \ln \frac{2n^2}{\delta}} \right)/{\alpha}$.
As in Blum et al., we can also fix $\epsilon$ and bound the size of the database.  $\HH$ is $(\epsilon, \delta)$-useful when 
\[
N \geq \frac{16 \hght^{\frac{3}{2}}  \ln \frac{2n^2}{\delta}}{\alpha \epsilon}
= O\left( \frac{ \log^{\frac{3}{2}} n \; \left( \log n + \log \frac{1}{\delta} \right) } { \alpha \epsilon} \right)
\]

In comparison, the technique of Blum et al. is $(\epsilon, \delta)$-useful for range queries when 
\[
N \geq O\left( \frac{\log n \left( \log \log n + \log \frac{1}{\epsilon \delta}\right) } { \alpha \epsilon^3}  \right)
\]
Both techniques scale at most poly-logarithmically with the size of the domain.  However, the $\HH$ scales better with $\epsilon$, achieving the same utility guarantee with a database that is smaller by a factor of $O(1/\epsilon^2)$. 

The above comparison reveals a distinction between the techniques: for $\HH_q$ the bound on absolute error is independent of database size, i.e., it only depends on $\epsilon$, $\alpha$, and the size of range.  However, for the Blum et al. approach, the absolute error increases with the size of the database at a rate of $O(N^{2/3})$.

%
%
%
%
%
%
%
%
%
%
%
%
%
%


\end{document}